\documentclass[prd,twocolumn,superscriptaddress,preprintnumbers,balancelastpage,nofootinbib]{revtex4-1}

\usepackage{graphicx}  % needed for figures
\usepackage{dcolumn}   % needed for some tables
\usepackage{bm}        % for math
\usepackage{amssymb, amsmath}
\usepackage{slashed}   % for math
\usepackage[usenames,dvipsnames,svgnames,table]{xcolor}
\usepackage{comment}
\usepackage[utf8]{inputenc}

\def\lsim{\mathrel{\rlap{\lower4pt\hbox{\hskip1pt$\sim$}}
    \raise1pt\hbox{$<$}}}         %less than or approx. symbol
\def\gsim{\mathrel{\rlap{\lower4pt\hbox{\hskip1pt$\sim$}}
    \raise1pt\hbox{$>$}}}         %greater than or approx. symbol

\newcommand{\eq}[1]{Eq.\,(\ref{#1})}
\newcommand{\eqsand}[2]{Eqs.\,(\ref{#1}) and (\ref{#2})}
\newcommand{\eqsto}[2]{Eqs.\,(\ref{#1}) to (\ref{#2})}

\renewcommand{\thefootnote}{\fnsymbol{footnote}}
\def\beq#1\eeq{\begin{align}#1\end{align}}
\newcommand{\ov}{\overline}

\newcommand{\eg}{{\em e.g.}}
\newcommand{\ie}{{\em i.e.}}

\newcommand{\no}{\nonumber}
\newcommand{\imag}{\mbox{Im}\,}
\newcommand{\BR}{\mbox{BR}}

\renewcommand{\arraystretch}{1.3}

\usepackage{mathtools}
\newcommand{\wigeckinv}[2]{$\begin{bsmallmatrix}
			\mathbf{#1} \\
			#2
		\end{bsmallmatrix}$}

\RequirePackage{xspace}
\def\Bbar    {\kern 0.18em\overline{\kern -0.18em B}{}\xspace}

\def\Kbar    {\kern 0.18em\overline{\kern -0.18em K}{}\xspace}

\definecolor{BlueViolet}{rgb}{0.2, 0.00, 0.7}
\definecolor{Blue}{rgb}{0.15, 0.00, 0.9}
\definecolor{light_blue}{rgb}{0.15, 0.35, 0.95}
\definecolor{kitgreen}{rgb}{0, 
0.58823 %150/255
, 0.50980 %130/255
}

\usepackage[%dvipdfmx,
colorlinks=true, linkcolor=kitgreen,citecolor=kitgreen,urlcolor=kitgreen]{hyperref}

\begin{document}
\widetext
\preprint{KEK--TH--2643,  TTP24--029,  KA--TP--15--2024}

\title{$\bf{\rm{SU}(3)_F}$ sum rules for CP asymmetries of $D_{(s)}$ decays }

\author{Syuhei Iguro}
\email{igurosyuhei@gmail.com}
\affiliation{Institute for Theoretical Particle Physics (TTP), Karlsruhe Institute of Technology (KIT),
Wolfgang-Gaede-Str.\,1, 76131 Karlsruhe, Germany}
\affiliation{ Institute for Astroparticle Physics (IAP),
Karlsruhe Institute of Technology (KIT), 
Hermann-von-Helmholtz-Platz 1, 76344 Eggenstein-Leopoldshafen, Germany}
\affiliation{Institute for Advanced Research (IAR), Nagoya University, Nagoya 464--8601, Japan}
\affiliation{Kobayashi-Maskawa Institute (KMI) for the Origin of Particles and the Universe,\\ Nagoya University, Nagoya 464--8602, Japan}
\affiliation{KEK Theory Center, IPNS, KEK, Tsukuba 305--0801, Japan}
\author{Ulrich Nierste}
\email{ulrich.nierste@kit.edu}
\affiliation{Institute for Theoretical Particle Physics (TTP), Karlsruhe Institute of Technology (KIT),
Wolfgang-Gaede-Str.\,1, 76131 Karlsruhe, Germany}
\author{Emil Overduin}
\email{overduin.emil@gmail.com}
\affiliation{Institute for Theoretical Particle Physics (TTP), Karlsruhe Institute of Technology (KIT),
Wolfgang-Gaede-Str.\,1, 76131 Karlsruhe, Germany}
\affiliation{Institute for Theoretical Physics (ITP), Karlsruhe Institute of Technology (KIT),
Wolfgang-Gaede-Str.\,1, 76131 Karlsruhe, Germany}
\author{Maurice Sch\"{u}\ss ler}
\email{maurice.schuessler@gmail.com}
\affiliation{Institute for Theoretical Particle Physics (TTP), Karlsruhe Institute of Technology (KIT),
Wolfgang-Gaede-Str.\,1, 76131 Karlsruhe, Germany}
\affiliation{Institute for Theoretical Physics (ITP), Karlsruhe Institute of Technology (KIT),
Wolfgang-Gaede-Str.\,1, 76131 Karlsruhe, Germany}

%%%%%%%%%%%%%%%%%%%%%%%%%%%%%%%%%%%%%%%%
\begin{abstract}
\noindent
Charge-parity (CP) asymmetries in charm decays are extremely suppressed in the Standard Model and may well be dominated by new-physics contributions.
The LHCb collaboration reported the results of direct CP asymmetry measurements in  $D^0\to K^+ K^-$ and $D^0\to \pi^+\pi^-$ decays with unprecedented accuracy: $a_{\rm{CP}}(K^+ K^-)=(7.7\pm5.7)\times 10^{-4}$ and $a_{\rm{CP}}(\pi^+\pi^-)=(23.2\pm6.1)\times 10^{-4}$, with the latter quantity inferred from the precise measurement of $\Delta a_{\rm{CP}} =\, a_{\rm{CP}}(K^+ K^-)
  -a_{\rm{CP}}(\pi^+\pi^-) 
  =\, (-15.7\pm2.9)\times 10^{-4}$. When interpreted within the Standard Model, these values 
  indicate a breakdown of the approximate $U$-spin symmetry of QCD. If, however, this symmetry holds 
  and the data stem from new physics, other CP asymmetries should be enhanced as well.
We derive CP asymmetry sum rules based on $\rm{SU}(3)$ flavor symmetry for $D$ meson decays into a pair of pseudoscalar mesons as well as a pair of a pseudoscalar and a vector meson 
for two generic scenarios, with $\Delta U=0$ and $|\Delta U|=1$ interactions, respectively.
The correlations implied by the sum rules can be used to check the consistency between different measurements and to discriminate between these scenarios with future data. For instance, we find 
$a_{\mathrm{CP}}(\pi^{+}K^{* 0}) + a_{\mathrm{CP}}(K^{+}\overline{K}^{* 0}) = 0$ for $\Delta U=0$
new physics and the opposite relative sign for the $|\Delta U|=1$ case. One sum rule, connecting four decay modes, holds in both scenarios. We further extend our sum rules to certain differences of CP asymmetries from which the $D$ production asymmetries drop out.
\\
---------------------------------------------------------------------------------------------------------------------------------
\end{abstract}
%%%%%%%%%%%%%%%%%%%%%%%%%%%%%%%%%%%%%%%%

\maketitle
\renewcommand{\thefootnote}{\#\arabic{footnote}}
\setcounter{footnote}{0}

%%%%%%%%%%%%%%%%%%%%%%%%%%%%%%%%%%%%
\section{Introduction}
\label{Sec:introduction}
%%%%%%%%%%%%%%%%%%%%%%%%%%%%%%%%%%%%

In 2019 the LHCb collaboration reported the discovery of charm CP violation (CPV) in the measurement of the difference of two CP asymmetries stating \cite{LHCb:2019hro}
\begin{align}
  \Delta a_{\rm{CP}}^{2019}&=\, a_{\rm{CP}}(K^+ K^-)
  -a_{\rm{CP}}(\pi^+\pi^-) \nonumber\\
  &=\, (-15.7\pm2.9)\times 10^{-4}, \label{eq:dacp}
\end{align}
where $a_{\rm{CP}}(f)$ is the time-integrated direct CP asymmetry in $D^0\to f$. 
Strictly speaking, the measurement in \eq{eq:dacp} contains a small contribution from (the yet undiscovered) mixing-induced CP violation, because the average decay times of the $D^0 \to K^+ K^-$ and $D^0 \to \pi^+ \pi^-$ data samples are different.
In this paper we assign the measured values completely to direct CP asymmetry; subtracting the maximal experimentally allowed contribution from mixing-induced CP violation changes the central value of the direct CP asymmetry difference in \eq{eq:dacp} by as little as $+0.3\times 10^{-4}$ \cite{LHCb:2019hro}. 

In 2022 LHCb presented the corresponding measurement of the individual CP asymmetry $a_{\rm{CP}}(K^+ K^-)$ and combined it with \eq{eq:dacp} and previous measurements to find \cite{LHCb:2022lry}:
\begin{align}
  a_{\rm{CP}}(K^+ K^-)&=\, (7.7\pm5.7)\times 10^{-4}, \label{eq:k}\\
  a_{\rm{CP}}(\pi^+\pi^-)&=\, (23.2\pm6.1)\times 10^{-4}, \label{eq:pi}
\end{align}
with a correlation of $\rho=0.88$.

It is difficult to calculate Standard-Model (SM) predictions for the penguin amplitudes feeding \eqsand{eq:k}{eq:pi}.
However, the strong parametric suppression stemming from tiny off-diagonal elements of the Cabibbo-Kobayashi-Maskawa (CKM) matrix \cite{Cabibbo:1963yz,Kobayashi:1973fv} makes these CP asymmetries highly sensitive probes of new physics.
Virtual effects of multi-TeV mass heavy particles can easily dominate over the SM contribution, so that even SM predictions with ${\cal O}(100\%)$ uncertainty can constrain beyond-SM (BSM) models in a meaningful way.
A SM prediction based on QCD sum rules is \cite{Khodjamirian:2017zdu,Chala:2019fdb}
\begin{align}
|\Delta a_\text{CP}^{\text{SM}}|&=\, (2.4\pm1.2)\times 10^{-4},
\label{eq:sr}
\end{align}
which is smaller  than the measured value in \eq{eq:dacp} by a factor of more than 6.
While QCD sum rules are a sound, field-theoretic method with a plethora of successful predictions in $B$ physics, little is known about their applicability to charm physics.
The discrepancy between \eqsand{eq:dacp}{eq:sr} (as well as already the earlier, less significant measurement $\Delta a_\text{CP}=(-82\pm 21\pm 11)\times 10^{-4}$ \cite{LHCb:2011osy}) has stimulated many theory papers addressing either BSM physics
\cite{Grossman:2006jg,Isidori:2011qw,Grossman:2012eb,Giudice:2012qq,Hiller:2012wf,Altmannshofer:2012ur,Chen:2012usa,Keren-Zur:2012buf,Delaunay:2012cz,Dery:2019ysp,Chala:2019fdb,Calibbi:2019bay,Bause:2020obd,Buras:2021rdg,Bause:2022jes} or invoking a SM explanation in terms of an enhanced SM penguin amplitude.
The latter papers have postulated a QCD enhancement ad-hoc \cite{Brod:2012ud,Grossman:2019xcj} or by invoking unflavored resonances which are almost mass degenerate with the $D^0$ \cite{Soni:2019xko,Schacht:2021jaz}.

The approximate SU(3) flavor symmetry of QCD (SU(3)$_{\rm{F}}$) can be used to derive relations between various CP asymmetries and we expect to predict other non-vanishing direct CP asymmetries from the non-zero CP asymmetry in \eq{eq:pi}.
To this end we employ the subgroup of SU(3)$_{\rm{F}}$ corresponding to SU(2) rotations of the $U$-spin doublet $(s,d)^T$.
The $U$-spin symmetry breaking parameter is $(m_s-m_d)/\Lambda_{\rm QCD}$, so that one expects $U$-spin relations to hold up to corrections of order $30\%$.
SU(3)$_{\rm{F}}$ analyses of branching ratios (BR) and CP asymmetries can be found in Refs.~\cite{Pirtskhalava:2011va, Bhattacharya:2012ah, Feldmann:2012js, Bhattacharya:2012kq, Grossman:2012ry, Hiller:2012xm, Grossman:2013lya,Hiller:2013awa, Muller:2015lua, Muller:2015rna,  Gavrilova:2022hbx}.
However, within the SM one finds the sum rule 
\begin{align}
a_{\rm{CP}}(\pi^+\pi^-)=-a_{\rm{CP}}(K^+ K^-),
\label{Eq:Famous}
\end{align}
valid in the limit of exact $U$-spin symmetry.
As seen the sum rule predicts opposite signs of $a_{\rm{CP}}(\pi^+\pi^-)$ and $a_{\rm{CP}}(K^+ K^-)$, and deviates from the measurement by about 2.7$\sigma$ \cite{LHCb:2022lry}. 

Thus the experimental results in \eqsand{eq:k}{eq:pi} imply that
\begin{itemize}
\item[(i)] $U$-spin breaks down in charm CP asymmetries
\item[(ii)]  or the dominant contribution to at least one of the two CP asymmetries in \eqsand{eq:k}{eq:pi} stems from new physics (NP) \cite{Schacht:2022kuj,Bause:2022jes} 
\item[(iii)] or future measurements will find different values for $a_{\rm{CP}}(K^+ K^-)$ and/or $\Delta a_\text{CP}$; this possibility necessarily implies a shift of $a_{\rm{CP}}(K^+ K^-)$ by more than 2$\sigma$ with a change of the sign. (We do not consider the possibility that  $\Delta a_\text{CP}$ will change by far more than $5\sigma$ to comply with $a_{\rm{CP}}(K^+ K^-)>0$).  
\end{itemize}

Although the short-distance partonic level quark transitions can be evaluated perturbatively, the hadronic $D$ meson decays and its CPV parameters involve hadronic matrix elements such as $\langle K^+ K^-| (\overline{u} \gamma_\mu P_L s) (\overline{s} \gamma^\mu P_L c)| D^0\rangle$, which are not easily evaluated.

In this paper, we rely on the approximate $\rm{SU}(3)_{\rm{F}}$ symmetry of the QCD Lagrangian to correlate the amplitudes of different decay modes with the goal of discriminating between the three explanations listed above.
Previously, SU(3)$_{\rm{F}}$ sum rules for amplitudes, decay rates, and CP asymmetries of charmed meson decays in the SM were derived in Refs.\,\cite{Grossman:2012ry,Grossman:2013lya}.\footnote{See also Ref.~\cite{Nierste:2017cua} and references therein.} 
% Deriving the CP asymmetry sum rule involving the NP interaction is the goal of the paper.
The focus of this paper, however, is a BSM explanation of the apparent breakdown 
of U-spin symmetry seen in \eqsand{eq:k}{eq:pi}, considering concrete scenarios of generic
BSM Lagrangians involving U-spin breaking or U-spin conserving parameters. One of our scenarios resembles the SM case and we will compare our results with those of Ref.~\cite{Grossman:2013lya} where possible.
%The earlier work of Ref.\,\cite{Grossman:2013lya} also discusses CP violation in charmless B decays in the framework of $\rm{SU}(3)_{\rm{F}}$ within the SM.

For example, if (ii) is the correct explanation while U-spin holds, the pattern of \eqsand{eq:k}{eq:pi} will have imprints on other decay modes.
%Deriving the CP asymmetry sum rule involving the NP interaction is the goal of the paper.
Furthermore, the comparisons of CP asymmetries in $D_{(s)}$ decays to two pseudoscalar mesons with those in decays to a pseudoscalar/vector meson pair will give insight into the Dirac structure of the underlying BSM couplings.
We will derive SU$(3)_{\rm{F}}$ sum rules for both classes of decays.

It is worthwhile to mention that there are $\rm{SU}(3)_{\rm{F}}$ sum rules for $D$ decay rates which hold up to linear order in the $\rm{SU}(3)_{\rm{F}}$ breaking parameter \cite{Grossman:2012ry} but there is no sum rule for CP asymmetries in $D$ decays to this order
\cite{Grossman:2013lya}. In the SM CP asymmetries stem from the interference of the dominant tree amplitude with a CKM-suppressed penguin amplitude and one can only improve the predictions by including $\rm{SU}(3)_{\rm{F}}$ breaking in the tree amplitude while
staying in the $\rm{SU}(3)_{\rm{F}}$ limit for the penguin amplitude \cite{Muller:2015rna}.

The outline of this paper is as follows.
In Sec.\,\ref{Sec:frame}, we explain the setup and 
in Sec.\,\ref{Sec:CPASR} the CP asymmetry sum rules are derived.
In Sec.\,\ref{Sec:Extended}, we newly extend our sum rules to differences of CP asymmetries modeled after \eq{eq:dacp} in order to eliminate experimental production asymmetries.
We conclude in Sec.\,\ref{Sec:Conclusion}.

%%%%%%%%%%%%%%%%%%%%%%%%%%%%%%%%%%%%
\section{Framework}
\label{Sec:frame}
%%%%%%%%%%%%%%%%%%%%%%%%%%%%%%%%%%%%

%%%%%%%%%%%%%%%%%%%%%%%%%%%%%%%%%%%%%%%%%%%%%%%%%
\begin{figure}[t]
\begin{center}
\includegraphics[width=0.483\textwidth]{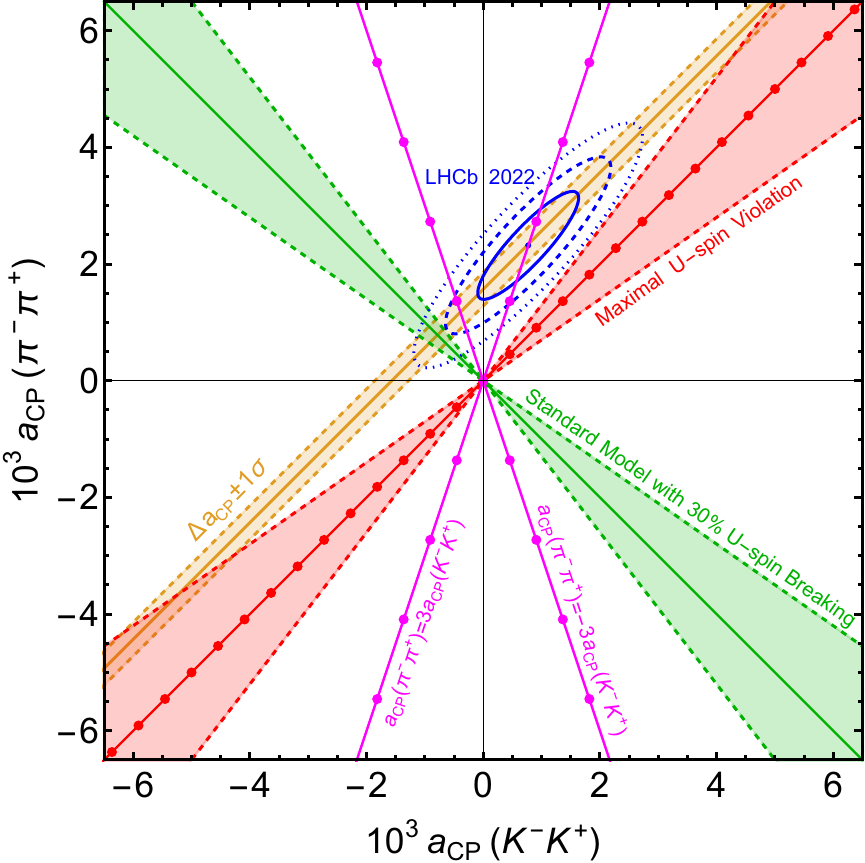}
\caption{
   \label{Fig:LHCb}
The experimental result and theory predictions on $a_\text{CP}(K^+K^-)$ vs $a_\text{CP}(\pi^+\pi^-)$ plane. 
See the text at the end of Sec.~\ref{Sec:frame} for details.
} 
\end{center}
\vspace{-1mm} \hrule
\end{figure}
%%%%%%%%%%%%%%%%%%%%%%%%%%%%%%%%%%%%%%%%%%%%%%%%

Within the SM the decays of interest are induced by the singly Cabibbo suppressed (SCS) charm decays $c\to u q\bar q$ with $q=d,s$ at tree-level and $q=u,d,s$ in the loop-induced penguin contribution.
The relevant $|\Delta C|=1$ effective Hamiltonian is given at the interaction scale ($\mu=m_c$) as
\begin{align}
    \mathcal{H}_{\rm{SM}}^{\rm{eff}}=&\frac{4G_F}{\sqrt{2}}\sum_{q=s,d} \lambda_{q} \biggl( C_1(\overline{q}^\alpha\gamma^\mu P_L c^\alpha)( \overline{u}^\beta\gamma_\mu P_L q^\beta)\nonumber\\
    &~~~~~~~~~~~~~~~~~~~~+C_2(\overline{q}^\alpha\gamma^\mu P_L c^\beta)( \overline{u}^\beta\gamma_\mu P_L q^\alpha)\biggl)\nonumber\\
    \equiv&\lambda_s h^s_{\rm{SM}} +\lambda_d h^d_{\rm{SM}},\label{eq:hsm}
\end{align}
where $\lambda_q=V^{\phantom{*}}_{uq}V_{cq}^*$  and $\alpha,\beta$ are color indices.
Defining $A_\pi=\langle \pi^+\pi^- | h^d_{\rm{SM}}-h^s_{\rm{SM}}|D^0\rangle$, $A_K=\langle K^+K^- | h^s_{\rm{SM}}-h^d_{\rm{SM}}|D^0\rangle$, $P_\pi=\langle \pi^+\pi^- | h^s_{\rm{SM}}|D^0\rangle$ and $P_K=\langle K^+K^- |h^d_{\rm{SM}}|D^0\rangle$, the decay amplitudes are expressed as 
\begin{align}
    \mathcal{A}(D^0\to \pi^+\pi^-)=&\langle \pi^+\pi^-|\mathcal{H}_{\rm{SM}}^{\rm{eff}}|D^0\rangle\nonumber\\
    =&\lambda_d A_\pi-\lambda_b P_\pi, \label{eq:apipi}\\
    \mathcal{A}(D^0\to K^+K^-)=&\lambda_s A_K-\lambda_b P_K, \label{eq:akk}
\end{align}
where the CKM unitarity relation $\lambda_d+\lambda_s+\lambda_b=0 $ is used. 
The $b$ quark is integrated out leading to penguin operators with tiny coefficients in $\mathcal{H}_{\rm{SM}}^{\rm{eff}}$ which come with $\lambda_b$ and are omitted in \eq{eq:hsm}.
These small terms contribute equally to $h^s_{\rm{SM}}$ and $h^d_{\rm{SM}}$.

From \eqsand{eq:apipi}{eq:akk} one notes that the meson pair is produced in a $U=1$ state in the limit $\lambda_b=0$.
It is straightforward to calculate the direct CP asymmetry 
\begin{align}
    a_\text{CP}\equiv\frac{|\mathcal{A}_{i\to f}|^2-|\mathcal{A}_{\overline{i}\to \overline{f}}|^2}{|\mathcal{A}_{i\to f}|^2+|\mathcal{A}_{\overline{i}\to \overline{f}}|^2},
\end{align}
and we obtain 
\begin{align}
    a_\text{CP}(\pi^+\pi^-)\simeq \,2\, \imag \frac{\lambda_b}{\lambda_d}
    \imag\left(\frac{P_\pi}{A_\pi}\right),\\
    a_\text{CP}(K^+K^-)\simeq \,2\,  \imag \frac{\lambda_b }{\lambda_s} \imag\left(\frac{P_K}{A_K}\right),
    \end{align}
where we neglected $\mathcal{O}(\lambda_b^2) $ terms.
Given that $\lambda_s= -\lambda_d+\mathcal{O}(\lambda_b) $ holds thanks to CKM unitarity, $U$-spin symmetry ensures $A_\pi=A_K$ and $P_\pi=P_K$, reproducing the famous CP asymmetry sum rule of Eq.\,(\ref{Eq:Famous}).

In reality $\rm{SU}(3)_{\rm{F}}$ is broken and the breaking effect is found to be $\simeq 30\%$ in the dominant decay amplitude $\propto \lambda_{d,s}$ in measurements of BRs \cite{Grossman:2012ry}. 
In this paper, we employ $\rm{SU}(3)_{\rm{F}}$ at the leading order and neglect $U$-spin symmetry violation from QCD in the matrix elements since the observed violation 
%which would point to the NP}
in \eqsand{eq:k}{eq:pi} is huge and its interpretation in terms of BSM physics does not need a better precision than the ${\cal O}(30\%)$ accuracy of the $\rm{SU}(3)_{\rm{F}}$ limit.
In the SM the penguin contributions  $P_\pi$ and $P_K$ are $\Delta U =0$ amplitudes. 
$\Delta U =0$ NP contributes to $a_{\rm{CP}}(\pi^+\pi^-)$ and $a_{\rm{CP}}(K^+ K^-)$ with opposite sign as the SM one, so that a NP explanation of \eqsand{eq:k}{eq:pi} requires a $|\Delta U| =1$ contribution. 
If such a contribution is observed in future measurements of other CP asymmetries, this will corroborate the NP interpretation.
To this end, we derive CP sum rules for decays in which non-zero CP asymmetries are not yet observed.

Generic NP four-quark $\Delta S=0$ interactions can be described by amending the effective SM Hamiltonian in \eq{eq:hsm} with
\begin{align}
\label{Eq:NP}
\Delta\mathcal{H}_{\rm{NP}}^{\rm{eff}}&=\frac{G_F}{\sqrt{2}}(\ov{u}\Gamma c)
 \left( a_u\,\ov{u}\Gamma u\, +\, a_d\,\ov{d}\Gamma d\, +\, a_s\, \ov{s}\Gamma s \right)
 \nonumber\\
&\equiv a_u \mathcal{O}^\prime_u +  a_d \mathcal{O}^\prime_d +a_s \mathcal{O}^\prime_s\,,
\end{align}
where $\Gamma$ represents an arbitrary Dirac structure. 
While several such terms with different Dirac structures could be present, our symmetry-based analyses will not be changed compared to the case in \eq{Eq:NP} with a single Dirac structure.
The same remark applies to the two possible color structures; color indices are not shown in \eq{Eq:NP}.

Returning to $D^0\to \pi^+\pi^-$ and $K^+K^-$, we set $a_u=0$ until the end of this section, because 
$a_u$ contributes only through penguin or annihilation diagrams to these decays which are likely to be smaller than tree-level NP effects involving $a_d$ or $a_s$.
The contributing amplitudes in the presence of NP effects, $\mathcal{H}_{\rm{NP}}^{\rm{eff}}=\mathcal{H}_{\rm{SM}}^{\rm{eff}}+\Delta\mathcal{H}_{\rm{NP}}^{\rm{eff}}$ are expressed as,
\begin{align}
    A^{\rm{NP}}(D^0\to \pi^+\pi^-)=&\,\lambda_d A_\pi+a_d\ \mathcal{Q}_d^\pi +a_s \mathcal{Q}_s^\pi,\\
    A^{\rm{NP}}(D^0\to K^+K^-)=&\,\lambda_s A_K+a_d\ \mathcal{Q}^K_d +a_s \mathcal{Q}_s^K,
\end{align}
where $\mathcal{Q}_q^M=\langle M\overline{M}|\mathcal{O}_q^\prime|D^0\rangle$ is defined. 
$\mathcal{Q}_s^\pi$ and $\mathcal{Q}_d^K$ involve $s$ and $d$ loops, respectively.
For instance, if we introduce NP which only couples to $d$ quarks, this corresponds to $a_d\neq 0$ and $a_s=0$.
We emphasize that as long as the involved hadronic matrix elements cannot be determined accurately, the $\Delta U=0$ NP contribution $\propto a_d+a_s$ cannot be disentangled from the SM contribution.
Similar to the SM case we obtain
{\small
\begin{align}
    a_\text{CP}(\pi^+\pi^-)&\simeq -2\, \imag \frac{a_d}{\lambda_d} 
    \,\imag\frac{\mathcal{Q}_d^\pi}{A}-2 \, \imag  \frac{a_s }{\lambda_d}\,\imag \frac{\mathcal{Q}_s^\pi}{A},\\
    a_\text{CP}(K^+K^-)&\simeq -2\, \imag \frac{a_d }{\lambda_s} 
     \,\imag \frac{\mathcal{Q}_d^K}{A}-2\, \imag \frac{ a_s }{\lambda_s}
     \,\imag\frac{\mathcal{Q}_s^K}{A},\label{eq:acpnp}
\end{align}}working to leading order in $\lambda_b$ and $a_{d,s}$ and using the $U$-spin symmetry which holds approximately, $A\equiv A_\pi=A_K$, in the SM part.
The maximal $U$-spin breaking in $\Delta a_\text{CP}$ corresponds to $\imag (a_d+a_s)=0$.
In this scenario $a_\text{CP}(\pi^+\pi^-)=a_\text{CP}(K^+K^-)$ holds, however, this relation also does not fit the recent data.

In Fig.~\ref{Fig:LHCb} we show the experimental status and theory predictions in the $a_\text{CP}(K^+K^-)$ vs $a_\text{CP}(\pi^+\pi^-)$ plane.
The 2019 LHCb result for $\Delta a_\text{CP}$ is shown in orange with $1\sigma$ uncertainty.
We show the latest LHCb result of $1,~2$, and $3\sigma$ in blue solid, dashed, and dotted ellipses. 
The $U$-spin limit,\,$a_\text{CP}(K^+K^-)=-a_\text{CP}(\pi^+\pi^-)$ as well as the limit of maximal $U$-spin violation,\,$a_\text{CP}(K^+K^-)=+a_\text{CP}(\pi^+\pi^-)$ are shown in light green and red, respectively.
To account for $U$-spin violating effects based on the ratio of BR$(D^0\to K^+K^-)$ and BR$(D^0\to \pi^+\pi^-)$ \cite{Grossman:2012ry}, we allow $30\%$ deviation from the $U$-spin limit corresponding to the green band. 
The magenta lines correspond to the reference  points, $a_\text{CP}(\pi^+\pi^-)=\pm3\,a_\text{CP}(K^+K^-)$. 
These reference points are motivated by the case 
$a_d\neq 0$ with $a_s=0$ and the observation that $\mathcal{Q}_d^K $ is color-suppressed w.r.t.\ $\mathcal{Q}_d^\pi $. 
The phase difference between $\mathcal{Q}_d^K/A $ and $\mathcal{Q}_d^\pi/A $ can be anything 
and the two signs in $a_\text{CP}(\pi^+\pi^-)=\pm3\,a_\text{CP}(K^+K^-)$ are the limiting cases if the color suppression is at its nominal value of $1/N_c=1/3$.

Assuming $\imag\left(\frac{\mathcal{Q}_d^\pi}{A}\right)=1$, the distance between any two points next to each other on the lines corresponds to $\Delta \imag (a_d)=0.05\times 10^{-3}$.
It is evident that $|\Delta U|=1$ \ie\,\,maximal $U$-spin violation cannot explain the data either while the central value of the recent LHCb data can be reproduced with $a_\text{CP}(\pi^+\pi^-)=+3\,a_\text{CP}(K^+K^-)$.

%%%%%%%%%%%%%%%%%%%%%%%%%%%%%%%%%%%%
\section{CP asymmetry sum rules}
\label{Sec:CPASR}
%%%%%%%%%%%%%%%%%%%%%%%%%%%%%%%%%%%%

In this section, we derive sum rules connecting new CP asymmetries, valid for $\Delta U=0$ and $|\Delta U|=1$ interactions, respectively.
The derivation is similar to that of the amplitude sum rule performed in Ref.\,\cite{Grossman:2012ry}, however, it is hard to find CP asymmetry sum rules in general, since CP asymmetries involve interference between two amplitudes, and thus the number of independent relations is smaller.
For demonstration, we start with a generic decay amplitude of 
\begin{align}
	\langle M_1 M_2 | \mathcal{H} | D \rangle = \lambda \, A^{(M_1M_2)} + a \, P^{(M_1M_2)}.
\end{align}
This leads to a CP asymmetry of 
\begin{align}
	a_\text{CP}(M_1M_2) \simeq~~~~~~~& \no\\
    + 2 \, \frac{ \imag \left( a \right) }{\lambda}&\, \frac{\imag{\left(A^{(M_1M_2)}\, P^{(M_1M_2)}{}^*\right)}}{|A^{(M_1M_2)}|^2},
\end{align}
where terms of order $a^2$ and higher are neglected, and the relative complex phase is put in $a$ while $\lambda$ is chosen real.
It is helpful to decompose the amplitude via the Wigner-Eckart theorem, which allows us to rewrite the amplitude in terms of Clebsch-Gordan (CG) coefficients \cite{deSwart:1963pdg,Kaeding:1995vq} and reduced matrix elements, to construct the desired CP asymmetry sum rules.
The relevant coefficients are summarized in Tab.\,\ref{table:WEI_PP} and Tabs.\,\ref{table:WEI_PV},\,\ref{table:WEI_PV2} for $D\to PP$ and $D\to PV$, where $P$ and $V$ stand for a pseudoscalar meson and vector meson, respectively.

Schematically, we decompose the amplitudes $A^{(M_1M_2)}$ and $P^{(M_1M_2)}$ as,
\begin{align}
	A^{(M_1M_2)}= \sum_{i=1}^n c_i(M_1M_2) x_i,\\
	P^{(M_1M_2)}{}^*  = \sum_{j=1}^m c_j^\prime(M_1M_2) y_j, \label{eq:cp}
\end{align} 
where $x_i,\,y_j$ are in general independent reduced matrix elements and $c_i(M_1M_2),~c^\prime_j(M_1M_2)$ correspond to the CG coefficients where the relevant entries are summarized in Appendix\,\ref{Sec:WEI}.
The product is expressed as 
\begin{align}
	\delta_\text{CP}(M_1M_2) &\equiv A^{(M_1M_2)}P^{(M_1M_2)}{}^* \nonumber\\
	&= \sum_{i=1}^n \sum_{j=1}^m c_i(M_1M_2) c^\prime_j(M_1M_2) x_i y_j \nonumber\\
	&= (c_1c^\prime_1,c_1c^\prime_2,..,c_n c^\prime_m) \cdot
	\begin{pmatrix}
		x_1 y_1 \\
		\vdots \\
		x_n y_m
	\end{pmatrix} \no \\
&	\equiv \mathbf{c}(M_1M_2)^T \cdot \mathbf{x} \, .
\end{align}
In this convention, CP asymmetries are expressed as
\begin{align} 
    &a_\text{CP}(M_1M_2)=\no\\ 
     &\hspace{10ex} \chi(M_1M_2)\, \imag \bigl(\delta_\text{CP}(M_1M_2) \bigl)\,\imag(a), 
    \label{eq:acp_decomp}
\end{align}
where $\chi(M_1M_2)= +2/(\lambda |A^{(M_1M_2)}|^2)$ is defined.
We can express $|A^{(M_1M_2)}|^2$ in terms of the decay rate $\Gamma$, which is experimentally found from the measured BR:
\begin{align}
\lambda^2 |A^{(M_1M_2)}|^2 &=\, \Gamma (D\to M_1 M_2) \no\\
&=\; 
      \BR(D\to M_1 M_2)/\tau_D, \label{eq:asq} 
\end{align}
where the different lifetimes $\tau_D$ for $D=D^0,D^+,D_s^+$ must be taken into account.
\eq{eq:asq} holds, because $|P^{(M_1M_2)}|$ is too small to have an effect on $\BR(D\to M_1 M_2)$.
The phase space factor is absorbed into the definition of $A^{(M_1M_2)}$; the phase space factors of different two-body $D$ decays are equal in the SU(3)$_{\rm{F}}$ symmetry limit and indeed do not differ much from each other. 
We have
\begin{align}
\chi(M_1M_2)&=\, \frac{2\, \lambda \tau_D}{\BR (D\to M_1 M_2)}, \label{eq:defchi}
\end{align}
and hence
\begin{align}
\imag \bigl( \delta_\text{CP}(M_1M_2) \bigl)& \,\lambda \,\imag(a) =\no \\
  ~~~~~~~~~~~~~~&a_\text{CP}(M_1M_2) \frac{\BR (D\to M_1 M_2)}{2 \, \tau_D}  
 .\label{eq:atode}
\end{align}
We will quote sum rules in terms of the $\delta_\text{CP}(M_1M_2)$'s; to relate these to the measured CP asymmetries, BRs, and lifetimes one must use \eq{eq:atode}. 
Since the sum rules are linear in the $\delta_\text{CP}(M_1M_2)$'s, the overall normalisation does not matter.
For this reason, also the theoretical parameter $\lambda\, \imag(a)$ drops out from the sum rules and cannot be determined.

Next, we consider a vector of all CP asymmetries
\begin{align}
    \mathbf{a}_\text{CP}^T= \left( a_{\mathrm{CP},1},a_{\mathrm{CP},2}...a_{\mathrm{CP},n}\right).
\end{align}
Then constructing sum rules is equivalent to finding a vector $\vec{v}$ orthogonal to $\mathbf{a}_\text{CP}$,
which satisfies
\begin{align}
    \mathbf{v}^T\cdot \mathbf{a}_\text{CP} =0. \label{eq:v}
\end{align}
If a sum rule involves only two modes, we can directly construct the $a_\text{CP}$ sum rule from the $\delta_\text{CP}$ sum rule, see, Appendix \ref{Sec:TMSR} for detail.

The general procedure discussed above can also be performed incorporating higher orders of $\rm{SU}(3)_{\rm{F}}$ breaking.
The cost is a larger number of involved reduced matrix elements.
We note that since the matrix $\mathbf{c}$ grows with the number of reduced matrix elements squared, it will be more difficult to find sum rules for CP asymmetries incorporating $\rm{SU}(3)_{\rm{F}}$ breaking effect.

To incorporate the generic interaction we consider the following general amplitude of the pseudoscalar decays,
\begin{align}
&	\langle M_1 M_2 | \mathcal{H} | D \rangle = \no\\
 &\hspace{5ex}\lambda_{SM} A^{(M_1M_2)} + a_0 P_{0}^{(M_1M_2)} + a_1 P_{1}^{(M_1M_2)},
\end{align}
where $P_{0}^{(M_1M_2)}$ and $P_{1}^{(M_1M_2)}$ correspond to the $\Delta U = 0$ and $|\Delta U| = 1$ contributions, respectively.
The SM $|\Delta U|=1$ amplitude with the CKM factor $\lambda_{SM}=\lambda_{d,s}$ is $A^{(M_1M_2)}$.
Both SM penguin and NP $\Delta U = 0$ contributions are contained in $P_{0}^{(M_1M_2)}$, while $P_{1}^{(M_1M_2)}$ stems solely from NP.
Keeping terms up to linear order in $a_0$ and $a_1$, the contributions to the CP asymmetry can be separated into two parts as
\begin{align} 
	a_\text{CP} &= a_\text{CP}^{\Delta U = 0} + a_\text{CP}^{\Delta U = 1} \nonumber\\
	&\simeq 2 \,\imag \frac{a_0}{\lambda_{SM}} \frac{\text{Im}(A\, P_0^*)}{|A|^2} + 2\, \imag \frac{a_1}{\lambda_{SM}} \, \frac{\text{Im}(A\, P_1^*)}{|A|^2} .
 \label{Eq:aCP_U01}
\end{align}
It is difficult to find vectors $\mathbf{v}$ satisfying \eq{eq:v} if both $a_0$ and $a_1$ are non-zero, because the CG coefficients $c_j^\prime$ in \eq{eq:cp} are different for $\Delta U = 0$ and $|\Delta U| = 1$ matrix elements in general.  

In the following two sub-sections (Secs.\,\ref{Sec:DACPSR_PP} and \ref{Sec:DACPSR_PV}) we present CP asymmetry sum rules for $D\to PP$ and $D\to PV$, respectively, considering the cases $a_1= 0$ and $a_0=0$. 
We adopt the $\rm{SU}(3)_{\rm{F}}$ limit for both the Hamiltonian and the meson states, \textit{e.g.} we identify $\eta$ and $\eta^\prime$ with octet and singlet states $\eta_8$ and $\eta_1$, respectively.
Specifically, we consider two scenarios, characterised by U-spin $U$ and isospin $I$: 
\begin{itemize} 
\item[I.] $\Delta U=\Delta I=0$: We assume $\Delta\mathcal{H}_{\rm{NP}}^{\rm{eff}}$ in \eq{Eq:NP} to be an SU(3) singlet, $a_0\propto a_u=a_d=a_s$ and $a_1=0$. This NP scenario mimics the SM penguin contribution, but with $a_0$ unrelated to $\lambda_b$. 
\item[II.]  $|\Delta U| = 1$ with $a_0=a_u=0$ and $a_1\propto a_s=-a_d$:   
 This NP scenario is motivated by a heavy new charged particle, such as a charged Higgs boson, though such a particle will also involve $\Delta U=0$ interactions (and effects on doubly Cabibbo suppressed (DCS) decays) as well.
 Also, a neutral particle with flavor changing neutral current (FCNC) $\bar u c$ coupling could produce this situation, if the coupling to up quarks is suppressed.
 LHC collider physics data place a lower bound of $\sim 22\;$TeV on the scale of four-quark interactions \cite{ATLAS:2017eqx,ATLAS:2023fod}.
 If the complex phase of the corresponding coupling is ${\cal O}(1)$ the CP asymmetries studied in this paper are more sensitive to effective $\bar u c \bar q q$ couplings than collider searches.
  \end{itemize}

%%%%%%%%%%%%%%%%%%%%%%%%%%%%%%%%%%%%
\subsection{\texorpdfstring{$D\to PP$}{D->PP}}
\label{Sec:DACPSR_PP}
%%%%%%%%%%%%%%%%%%%%%%%%%%%%%%%%%%%%
\subsubsection{$\Delta U$ = 0 New Physics}

First, we present CP asymmetry sum rules which hold for $\Delta U=0$ interactions ($a_0\neq0,\,a_1=0$).
%%%%%%%%%%%%%%%%%%%%%%%%%%%%%%%%%
The following two sum rules are well known and are found by a naive interchange of $d$ and $s$ quarks.
\begin{align}
a_\text{CP}^{\Delta U =0}(K^-K^+) + a_\text{CP}^{\Delta U =0}(\pi^-\pi^+) = 0, \label{Eq:PPU0_1}\\
a_\text{CP}^{\Delta U =0}(K^0\pi^+) + a_\text{CP}^{\Delta U =0}(\overline{K}^0K^+) = 0 .\label{Eq:PPU0_2}
\end{align}
To clarify our notation, remember that SCS decays do not change the strangeness $S$, so that in this section all $a_\text{CP} (M_1^0M_2^+)$'s in which the final state $M_1^0M_2^+$ has $S=1$ (like those in \eq{Eq:PPU0_2}) stem from $D_s^+ \to M_1M_2^+ $.

The first sum rule Eq.\,(\ref{Eq:PPU0_1}) is the same as Eq.\,(\ref{Eq:Famous}) and found to be violated by the latest measurements.
The experimental data lead to 
\begin{align}
	a_\text{CP}(K^-K^+) + a_\text{CP}(\pi^-\pi^+) = (30.9\pm11.4) \times 10^{-4} \, ,
\end{align}
which deviates from the $\Delta U =0$ sum rule by more than $2\sigma$, as seen in Fig.\,\ref{Fig:LHCb}.
This is already an interesting hint that there may be more contributions beyond the $\Delta U = 0$ penguin interaction.
We also have the following sum rules,
\begin{align} 
&a_\text{CP}^{\Delta U =0}(\pi^0\pi^+) = 0,\label{Eq:PPU0_3}\\
&\delta_\text{CP}^{\Delta U = 0}(\eta_8\eta_8) +  \delta_\text{CP}^{\Delta U = 0}(\pi^0\pi^0) + 2\, \delta_\text{CP}^{\Delta U =0}(\eta_8\pi^0) = 0,\label{Eq:PPU0_4}\\
&\delta_\text{CP}^{\Delta U =0}(\eta_8K^+) + \delta_\text{CP}^{\Delta U =0}(\eta_8\pi^+) + \delta_\text{CP}^{\Delta U =0}(\pi^0K^+) = 0,\label{Eq:PPU0_5}\\
&3\,\delta_\text{CP}^{\Delta U =0}(\eta_8K^+) -3\,\delta_\text{CP}^{\Delta U =0}(\pi^0K^+) +\delta_\text{CP}^{\Delta U =0}(K^0\pi^+) = 0\label{Eq:PPU0_6}.
\end{align}
\eq{Eq:PPU0_3} is, of course, a well-known null test of the SM, which is not violated if the NP contribution is pure $\Delta I=0$ as in the considered SU(3) singlet NP scenario. $\eta_8$ is the octet $\eta$ meson. 
The physical $\eta$ meson is dominantly $\eta_8$ plus a smaller admixture of the singlet state $\eta_1$.
The associated mixing angle vanishes in the limit of exact SU(3)$_{\rm{F}}$ symmetry; since we neglect SU(3)$_{\rm{F}}$ breaking, the sum rules quoted in this paper can be used with the replacement $\eta_8\to\eta$. 
In \eqsand{Eq:PPU0_1}{Eq:PPU0_2} as well as
\eqsand{Eq:PPU0_4}{Eq:PPU0_5} we confirm the sum rules derived in Ref.~\cite{Grossman:2013lya} for the SM.
In this reference a $\eta-\eta^\prime$ mixing angle is included and the analogues of \eqsand{Eq:PPU0_4}{Eq:PPU0_5} involve additional decay modes, with final states containing $\eta^\prime$ mesons.
It is not possible to do this for our new sum rule in \eq{Eq:PPU0_6}.
This would require that the  involved
$D\to P\eta_8$ decay amplitudes fulfill the same sum rule as their
$D\to P\eta_1$ counterparts, which is not the case for \eq{Eq:PPU0_6}.  We list the sum rules for $D\to P \eta_1$ decays in Appendix~\ref{Sec:SiSuR}.
Adding the $\eta-\eta^\prime$ mixing angle to the sum rules in \eqsand{Eq:PPU0_4}{Eq:PPU0_5} includes one source of SU(3)$_{\rm{F}}$ breaking,
but this can be omitted as long as other  SU(3)$_{\rm{F}}$ breaking terms are omitted as well.

We emphasize that any linear combination of the above sum rules holds as well.
With \eq{eq:atode} one finds the sum rules for the CP asymmetries from the ones quoted for the $\delta_\text{CP}$'s, with the overall factor $\lambda_{SM} \imag (a_0)$ dropping out.
If in future measurements these sum rules are violated significantly beyond the nominal $\sim 30\%$ U-spin breaking, this will be evidence of $|\Delta U|=1$ NP.

%%%%%%%%%%%%%%%%%%%%%%%%%%%%%%%%%%%%%%%%%%%%%%%%%%%%%%%%%%%%%%%%%%%%%%%%%
\subsubsection{$|\Delta U|$ = 1 New Physics}
%%%%%%%%%%%%%%%%%%%%%%%%%%%%%%%%%%%%%%%%%%%%%%%%%%%%%%%%%%%%%%%%%%%%%%%%%

Next, we consider CP asymmetry sum rules for $a_0=0$ and $a_1 \neq 0$.
There are two $|\Delta U| = 1$ contributions, one from the SM amplitude $\lambda_{SM}\,A^{(M_1M_2)}$ and one from the NP amplitude $a_1\,P_{1}^{(M_1M_2)}$ carrying a different CP phase. 
Here we can again use the described procedure in Appendix\,\ref{Sec:TMSR} to find the two-mode sum rules.
This time four sum rules exist which contain two decay modes each:
 \begin{align}
 a_\text{CP}^{\Delta U =1}(K^-K^+) - a_\text{CP}^{\Delta U = 1}(\pi^-\pi^+) = 0, \label{Eq:PPU1_1}\\
 a_\text{CP}^{\Delta U =1}(K^0\pi^+) - a_\text{CP}^{\Delta U =1}(\overline{K}^0K^+) = 0,\label{Eq:PPU1_2}\\
 a_\text{CP}^{\Delta U =1}(\eta_8\eta_8) - a_\text{CP}^{\Delta U =1}(\pi^0\pi^0) = 0, \label{Eq:PPU1_3}\\
 a_\text{CP}^{\Delta U =1}(\eta_8\eta_8) - a_\text{CP}^{\Delta U =1}(\eta_8\pi^0) = 0.\label{Eq:PPU1_4}
 \end{align}
Additionally we find two sum rules connecting four decay modes. They are given by
\begin{align}
&\delta_\text{CP}^{\Delta U =1}(\eta_8K^+) -\delta_\text{CP}^{\Delta U =1}(\pi^0\pi^+) -\delta_\text{CP}^{\Delta U =1}(\eta_8\pi^+) \nonumber\\
&+\delta_\text{CP}^{\Delta U =1}(\pi^0K^+) = 0, \label{Eq:PPU1_5}\\
&6\,\delta_\text{CP}^{\Delta U =1}(\pi^0\pi^+) - 3\,\delta_\text{CP}^{\Delta U =1}(\eta_8K^+) +3\,\delta_\text{CP}^{\Delta U =1}(\pi^0K^+)\nonumber\\
&-\delta_\text{CP}^{\Delta U =1}(K^0\pi^+) = 0.\label{Eq:PPU1_6}
\end{align}
The sum rule in Eq.\,(\ref{Eq:PPU1_1}) cannot explain the LHCb measurement of,
\begin{align} \label{eq:testsum}
	 \Delta a_\text{CP} &= a_\text{CP}(K^-K^+) - a_\text{CP}(\pi^-\pi^+)\nonumber\\
  &=(-15.7\pm2.9) \times 10^{-4} \, ,
\end{align}
because $|\Delta U|=1$ contributions drop out from $\Delta a_\text{CP}$, their sole effect is to 
shift $a_\text{CP}(K^-K^+)$ and $a_\text{CP}(\pi^-\pi^+)$ into the region compatible with the 
U-spin symmetric matrix elements.

Interestingly, the sum rule in Eq.\,(\ref{Eq:PPU1_6}) holds in both of our two scenarios;
for the $\Delta U=0$ case it is constructed as $6\times$\,Eq.\,(\ref{Eq:PPU0_3})\,$-$\,Eq.\,(\ref{Eq:PPU0_6}).
While current experimental data do not allow us to verify this sum rule due to large experimental uncertainties, it will give useful insight into the quality of SU(3) symmetry of the hadronic matrix elements in the future: In case that data will comply well with Eq.\,(\ref{Eq:PPU1_6}), one will 
gain confidence in the SU(3) method and use the other sum rules to discriminate between the scenarios. 
Note that a future establishment of $a_\text{CP}(\pi^0\pi^+)\neq 0$ will establish isospin-breaking NP and thereby falsify the SM and our scenario I; in our scenario II then at least one other 
CP asymmetry entering Eq.\,(\ref{Eq:PPU1_6}) will be sizable because of the factor  6 in front of 
$\delta_\text{CP}^{\Delta U =1}(\pi^0\pi^+)$.

Above we separately derived the CP asymmetry sum rules for $|\Delta U|=1$ and $\Delta U=0$ interactions assuming specific $a_{d,s,u}$ combinations for each.
As it is seen in Fig.\,\ref{Fig:LHCb}, neither single $|\Delta U|=1$ nor $\Delta U=0$ interactions can fully address the current data.
Current data prefer the ratio $a_s:a_d=1:-3$ which leads to 
\begin{align}
(a_s+a_d):(a_s-a_d)=-1:2.
\label{eq:ratio_U01}
\end{align}
Generally, one can combine the CP-asymmetry sum rules in this proportion and confront these weighted
sum rules with data. However, one should keep in mind that $a_u$ enters our scenarios differently: Modifying 
scenario I by choosing $a_u=0$ to comply with scenario II will still be a $\Delta U=0$ scenario, but now with isospin breaking, invalidating  \eq{Eq:PPU0_3}. Then two possibilities must be considered: If 
$a_\text{CP} (\pi^0\pi^+)\neq 0$ is measured, this will directly establish NP with $a_u\neq a_d$. 
Yet if $a_\text{CP} (\pi^0\pi^+)$ is measured to be compatible with zero, this means that either 
$\imag a_u\approx \imag a_d$ or that the strong phase between the SM tree amplitude and the NP amplitude 
$\propto a_u-a_d$ is small, see \eq{eq:acpnp}. In the latter case the effect of $a_u-a_d\neq 0$ drops out
from the $\Delta U=0$ sum rules and the above-mentioned weighted sum rules are meaningful.

%%%%%%%%%%%%%%%%%%%%%%%%%%%%%%%%%%%%
\subsection{\texorpdfstring{$D\to PV$}{D->PV}}
\label{Sec:DACPSR_PV}
%%%%%%%%%%%%%%%%%%%%%%%%%%%%%%%%%%%%

Next, we consider the CP sum rule for $D \to PV$ decays valid for the $\Delta U=0$ 
interactions of our scenario I. 
Considering the penguin operators for $\rm{SU}(3)$, the following sum rules hold,
\begin{align}
    &a^{\Delta U = 0}_{\mathrm{CP}}(K^{0}\overline{K}^{* 0}) + a^{\Delta U = 0}_{\mathrm{CP}}(\overline{K}^{0}K^{* 0}) = 0,\label{Eq:PVU0_1}\\
    &a^{\Delta U = 0}_{\mathrm{CP}}(K^{-}K^{* +}) + a^{\Delta U = 0}_{\mathrm{CP}}(\pi^{-}\rho^{+}) = 0,\label{Eq:PVU0_2}\\
    &a^{\Delta U = 0}_{\mathrm{CP}}(K^{+}K^{* -}) + a^{\Delta U = 0}_{\mathrm{CP}}(\pi^{+}\rho^{-}) = 0,\label{Eq:PVU0_3}\\
    &a^{\Delta U = 0}_{\mathrm{CP}}(K^{0}\rho^{+}) + a^{\Delta U = 0}_{\mathrm{CP}}(\overline{K}^{0}K^{* +}) = 0,\label{Eq:PVU0_4}\\
    &a^{\Delta U = 0}_{\mathrm{CP}}(\pi^{+}K^{* 0}) + a^{\Delta U = 0}_{\mathrm{CP}}(K^{+}\overline{K}^{* 0}) = 0,\label{Eq:PVU0_5}   \\
    &\delta^{\Delta U = 0}_{\mathrm{CP}}(\eta_{8}\omega_{8}) + \delta^{\Delta U = 0}_{\mathrm{CP}}(\eta_{8}\rho^{0}) + \delta^{\Delta U = 0}_{\mathrm{CP}}(\pi^{0}\omega_{8})\nonumber\\
    & + \delta^{\Delta U = 0}_{\mathrm{CP}}(\pi^{0}\rho^{0}) = 0,\label{Eq:PVU0_6}\\
    &\delta^{\Delta U = 0}_{\mathrm{CP}}(\pi^{0}K^{* +}) + \delta^{\Delta U = 0}_{\mathrm{CP}}(\eta_{8}K^{* +}) + \delta^{\Delta U = 0}_{\mathrm{CP}}(\pi^{0}\rho^{+}) \nonumber\\
    &+ \delta^{\Delta U = 0}_{\mathrm{CP}}(\eta_{8}\rho^{+})= 0,\label{Eq:PVU0_7}\end{align}\begin{align}
    &\delta^{\Delta U = 0}_{\mathrm{CP}}(K^{+}\rho^{0}) + \delta^{\Delta U = 0}_{\mathrm{CP}}(K^{+}\omega_{8}) + \delta^{\Delta U = 0}_{\mathrm{CP}}(\pi^{+}\rho^{0})\nonumber\\
    & + \delta^{\Delta U = 0}_{\mathrm{CP}}(\pi^{+}\omega_{8})= 0,\label{Eq:PVU0_8}\\
    &6 \delta^{\Delta U = 0}_{\mathrm{CP}}(\eta_{8}\rho^{0}) - 6\delta^{\Delta U = 0}_{\mathrm{CP}}(\pi^{0}\omega_{8}) - 5\delta^{\Delta U = 0}_{\mathrm{CP}}(\eta_{8}\rho^{+})\nonumber\\
    &- 3\delta^{\Delta U = 0}_{\mathrm{CP}}(\pi^{0}\rho^{+}) - \delta^{\Delta U = 0}_{\mathrm{CP}}(\overline{K}^{0}K^{* +})+\delta^{\Delta U = 0}_{\mathrm{CP}}(K^{+}\overline{K}^{* 0})\nonumber\\
    & + 3\delta^{\Delta U = 0}_{\mathrm{CP}}(\pi^{+}\omega_{8}) +    \delta^{\Delta U = 0}_{\mathrm{CP}}(\pi^{+}\rho^{0}) -     2\delta^{\Delta U = 0}_{\mathrm{CP}}(\eta_{8}K^{* +}) \nonumber\\
    &- 2\delta^{\Delta U = 0}_{\mathrm{CP}}(K^{+}\rho^{0}) = 0.\label{Eq:PVU0_9}
\end{align}
The first five sum rules, Eqs.\,(\ref{Eq:PVU0_1}-\ref{Eq:PVU0_5}) only involve 
two modes each, while the remaining three rules, Eqs.\,(\ref{Eq:PVU0_6}-\ref{Eq:PVU0_8}) relate four decay modes to each other.
The sum rules \eqsto{Eq:PVU0_1}{Eq:PVU0_8} confirm the earlier findings of Ref.~\cite{Grossman:2013lya}.
Our remark after \eq{Eq:PPU0_6} concerning the $\eta-\eta^\prime$ mixing angle applies here as well.
The last sum rule Eqs.\,(\ref{Eq:PVU0_9}), involving ten decay modes, is new. Like \eq{Eq:PPU0_6} one cannot add mixing-angle effects to \eq{Eq:PVU0_9}, because there are no matching sum rules for $D\to P \omega_1$ or $D\to \eta_1 V$ decays, see Appendix~\ref{Sec:SiSuR}.
Also our $D\to PV$ sum rules are to be understood with the replacement $\eta_8\to\eta$.
Further note that \eq{Eq:PVU0_9} involves both charged and neutral decays. 

On the other hand, once we assume that NP enters via $|\Delta U| = 1$ operators, the sum rules for CP-asymmetries can be written as
\begin{align}
    &a^{\Delta U = 1}_{\mathrm{CP}}(\pi^{0}\rho^{0}) - a^{\Delta U = 1}_{\mathrm{CP}}(\eta_{8}\omega_{8}) = 0,\label{Eq:PVU1_1}\\
    &a^{\Delta U = 1}_{\mathrm{CP}}(\overline{K}^{0}K^{* 0}) - a^{\Delta U = 1}_{\mathrm{CP}}(K^{0}\overline{K}^{* 0}) = 0,\label{Eq:PVU1_2}\\
    &a^{\Delta U = 1}_{\mathrm{CP}}(\pi^{-}\rho^{+}) - a^{\Delta U = 1}_{\mathrm{CP}}(K^{-}K^{* +}) = 0,\label{Eq:PVU1_3}\\
    &a^{\Delta U = 1}_{\mathrm{CP}}(\pi^{+}\rho^{-}) - a^{\Delta U = 1}_{\mathrm{CP}}(K^{+}K^{* -}) = 0,\label{Eq:PVU1_4}\\
    &a^{\Delta U = 1}_{\mathrm{CP}}(K^{0}\rho^{+}) - a^{\Delta U = 1}_{\mathrm{CP}}(\overline{K}^{0}K^{* +}) = 0,\label{Eq:PVU1_5}\\
    &a^{\Delta U = 1}_{\mathrm{CP}}(\pi^{+}K^{* 0}) - a^{\Delta U = 1}_{\mathrm{CP}}(K^{+}\overline{K}^{* 0}) = 0,\label{Eq:PVU1_6}\\
    &\delta^{\Delta U = 1}_{\mathrm{CP}}(\pi^{0}K^{* +}) + \delta^{\Delta U = 1}_{\mathrm{CP}}(\eta_{8}K^{* +}) - \delta^{\Delta U = 1}_{\mathrm{CP}}(\pi^{0}\rho^{+})\nonumber\\
    &- \delta^{\Delta U = 1}_{\mathrm{CP}}(\eta_{8}\rho^{+})= 0,\label{Eq:PVU1_7}\\
    &\delta^{\Delta U = 1}_{\mathrm{CP}}(K^{+}\rho^{0}) + \delta^{\Delta U = 1}_{\mathrm{CP}}(K^{+}\omega_{8}) - \delta^{\Delta U = 1}_{\mathrm{CP}}(\pi^{+}\rho^{0})\nonumber\\
    &- \delta^{\Delta U = 1}_{\mathrm{CP}}(\pi^{+}\omega_{8})= 0\label{Eq:PVU1_8}.
\end{align}
The other sum rules can be obtained by multiplying the individual $\delta_{\mathrm{CP}}$ with the corresponding $\chi$, see \eq{eq:defchi}. 
In the case of the $D \rightarrow PV$ decays no sum rule holds for $\Delta U=0$ and $1$ simultaneously.

%%%%%%%%%%%%%%%%%%%%%%%%%%%%%%%%%%%%
\section{Extended Sum Rule}
\label{Sec:Extended}
%%%%%%%%%%%%%%%%%%%%%%%%%%%%%%%%%%%%

Since except for $a_\text{CP}(\pi^-\pi^+)$ all CP asymmetries \cite{PDG2022} are currently measured consistent with zero (see, table \ref{tab:CPasym_future} of Appendix \ref{Sec:Input}), it is difficult to test sum rules at the present stage.
However, in the future Belle II and LHCb will reduce uncertainties by a factor of $\sim 5-10$ compared to the current measurements \cite{PDG2022,Belle-II:2018jsg,LHCb:2022lry,LHCb:2018roe} and could find more hints of CP violation.

To facilitate these discoveries we will next define differences $\Delta a_{\mathrm{CP}}$ 
of CP asymmetries in such a way that experimental production and detection asymmetries drop out.
In the past such considerations led to the measurement of $\Delta a_{\rm CP}$ in \eq{eq:dacp}. 
Our new $\Delta a_{\mathrm{CP}}$ combine each SCS CP asymmetry with another one in a 
Cabibbo favored (CF) and DCS decays. Sizable NP contributions to CF decays are not possible and only contrived models can generate a CP asymmetry in DCS decays \cite{Bergmann:1999pm} (see also Ref.~\cite{Bigi:1994aw}).
Therefore the $\Delta a_{\mathrm{CP}}$'s obey the same sum rules as the corresponding SCS CP asymmetry and will provide another important cross check.
We find 
\begin{align} 
 	\Delta a_\text{CP,1}(D_s^+) &= a_\text{CP}(K^0\pi^+) - a_\text{CP,CF}(\overline{K}^0K^+),\label{Eq:newdelta1}\\
	\Delta a_\text{CP,2}(D_s^+) &= a_\text{CP}( K^0\pi^+) - a_\text{CP,DCS}( K^0K^+),\\
	\Delta a_\text{CP,3}(D^+) &= a_\text{CP}(\overline{K}^0K^+) - a_\text{CP,CF}(\overline{K}^0\pi^+), \\
	\Delta a_\text{CP,4}(D^+) &= a_\text{CP}(\overline{K}^0K^+) - a_\text{CP,DCS}( K^0\pi^+),\label{Eq:newdelta2}
\end{align}
for $D\to PP$.
In reality, one does not observe a $K^0$ or $\bar K{}^0$, but a pair of two pions with the invariant mass of a kaon, i.e.\ a final state which approximately corresponds to a $K_S$. One must therefore subtract the effect of kaon CP violation from the data \cite{Grossman:2011zk}. This feature also leads to 
an interference in the CF and DCS decays, for example, the CP asymmetries for $D_s^+\rightarrow K_S K^+$ are non-zero.
The resulting CP asymmetries are proportional to the imaginary part of the ratio $V^*_{cd}V^{\phantom{*}}_{us}/V^*_{cs}V^{\phantom{*}}_{ud}$ in the SM and furthermore unlikely to be large even in the presence of NP \cite{Bergmann:1999pm} and thus negligible compared to the SCS CP asymmetries of interest.

As a result, these sum rules turn out to be very powerful because within the SM and SCS NP scenarios, the CP asymmetries for CF and DCS decays are highly suppressed.
Thus these differences essentially coincide with the CP asymmetries in the SCS decays of $D^+_s\rightarrow K^0\pi^+$ and $D^+ \rightarrow \overline{K}^0K^+$.
As said, the differences $\Delta a_\text{CP}$ in Eqs.\,(\ref{Eq:newdelta1})$-$(\ref{Eq:newdelta2}) are only taken for experimental reasons to eliminate the production asymmetries of $D^+$ and $D_s^+$, which can fake CP asymmetries. Thus we expect that the $\Delta a_{\rm CP,j}$'s in Eqs.\,(\ref{Eq:newdelta1})$-$(\ref{Eq:newdelta2}) can be measured more precisely than the single CP asymmetries.
But DCS decays might pose additional challenges since the amplitudes are further CKM suppressed and hence these decays are more difficult to access experimentally, so that 
 $\Delta a_{\text{CP},1}$ and $\Delta a_{\text{CP},3}$
 with CF-decays might be easier to measure. 

Similarly we can construct further $\Delta a_{\rm{CP}}$ observables for $D\to PV$ decays as
\begin{align}
            & \Delta a_{\mathrm{CP},5}(D^+_s) = a_{\mathrm{CP}}(K^{* 0} \pi^+) - a_{\mathrm{CP,DCS}}(K^0 K^{* +})  ,\\
            & \Delta a_{\mathrm{CP},6}(D^+_s) = a_{\mathrm{CP}}(K^{* 0} \pi^+) - a_{\mathrm{CP,DCS}}(K^{* 0} K^{+})   ,\\
            & \Delta a_{\mathrm{CP},7}(D^+_s) = a_{\mathrm{CP}}(K^0 \rho^+) - a_{\mathrm{CP,DCS}}(K^0 K^{* +})   ,\\
            & \Delta a_{\mathrm{CP},8}(D^+_s) = a_{\mathrm{CP}}(K^0 \rho^+) - a_{\mathrm{CP,DCS}}(K^{* 0} K^{+})  ,\\
            & \Delta a_{\mathrm{CP},9}(D^+) = a_{\mathrm{CP}}(\overline{K}^{* 0} K^+) - a_{\mathrm{CP,CF}}(\overline{K}^{* 0} \pi^{+})   ,\\
            & \Delta a_{\mathrm{CP},10}(D^+) = a_{\mathrm{CP}}(\overline{K}^{* 0} K^+) - a_{\mathrm{CP,DCS}}(K^{0} \rho^{+}),\\
            & \Delta a_{\mathrm{CP},11}(D^+) = a_{\mathrm{CP}}(\overline{K}^{0} K^{* +}) - a_{\mathrm{CP,CF}}(\overline{K}^{* 0} \pi^{+})   ,\\
            & \Delta a_{\mathrm{CP},12}(D^+) = a_{\mathrm{CP}}(\overline{K}^{0} K^{* +}) - a_{\mathrm{CP,DCS}}(K^{0} \rho^{+})   ,\\
            & \Delta a_{\mathrm{CP},13}(D^+) = a_{\mathrm{CP}}(\overline{K}^{* 0} K^+) {\small{-}} a_{\mathrm{CP,DCS}}(K^{* 0} \pi^{+})  ,\\
            & \Delta a_{\mathrm{CP},14}(D^+) = a_{\mathrm{CP}}(\overline{K}^{* 0} K^+) - a_{\mathrm{CP,CF}}(\overline{K}^{0} \rho^{+})  ,\\
            & \Delta a_{\mathrm{CP},15}(D^+) = a_{\mathrm{CP}}(\overline{K}^{0} K^{* +}) {\small{-}} a_{\mathrm{CP,DCS}}(K^{* 0} \pi^{+}) ,\\
            & \Delta a_{\mathrm{CP},16}(D^+) = a_{\mathrm{CP}}(\overline{K}^{0} K^{* +}) - a_{\mathrm{CP,CF}}(\overline{K}^{0} \rho^{+}) .
\end{align}
The comments made for $D\to PP$ decays also apply to $\Delta a_{\mathrm{CP},5-16}$ and potential CP asymmetries in the CF and DCS decays can be neglected. Note that in $\Delta a_{\mathrm{CP},6}$ and $\Delta a_{\mathrm{CP},9}$ also the $K^{* 0}\to K^+ \pi^-$ detection asymmetry cancels.

Precise measurements of these $\Delta a_{\rm{CP},j}$'s will serve to test the sum rules in Eq.\,(\ref{Eq:PPU1_2}),\,(\ref{Eq:PVU1_5}) and (\ref{Eq:PVU1_6}) for scenario II with $|\Delta U| = 1$ NP and $a_0 = 0$. As a result, in total, there are only three independent values of $\Delta a_{\mathrm{CP}}$, \eg \,$\Delta a_{\mathrm{CP},1}, \,\Delta a_{\mathrm{CP},5}$ and $\Delta a_{\mathrm{CP},7}$.
One finds
\begin{align}
	\Delta a_\text{CP,1} &= \Delta a_\text{CP,2} = \Delta a_\text{CP,3} = \Delta a_\text{CP,4} \\
    \Delta a_{\mathrm{CP},5} &= \Delta a_{\mathrm{CP},6} = \Delta a_{\mathrm{CP},9} = \Delta a_{\mathrm{CP},10} \nonumber\\&= \Delta a_{\mathrm{CP},13} = \Delta a_{\mathrm{CP},14},\\
    \Delta a_{\mathrm{CP},7} &= \Delta a_{\mathrm{CP},8} = \Delta a_{\mathrm{CP},11} = \Delta a_{\mathrm{CP},12} \nonumber\\&= \Delta a_{\mathrm{CP},15} = \Delta a_{\mathrm{CP},16}.
\end{align}
for vanishing CP asymmetries in CF and DCS decays. 
These relations can be useful to test the experimental consistency and the flavor structure of NP.

%%%%%%%%%%%%%%%%%%%%%%%%%%%%%%%%%%%%
\section{Conclusion}
\label{Sec:Conclusion}
%%%%%%%%%%%%%%%%%%%%%%%%%%%%%%%%%%%%
In this paper we revisited CP violation in hadronic two-body $D$ meson decays, motivated by the LHCb measurements of 
$a_{\rm CP} (D^0\to K^+ K^-)$ and $a_{\rm CP} (D^0\to \pi^+ \pi^-)$.
The data can only be accommodated within the Standard Model if the approximate 
$\rm{SU}(3)_{\rm{F}}$ symmetry of QCD fails for the penguin matrix elements entering these CP asymmetries
and furthermore a yet unknown mechanism enhances the size of the penguin matrix elements in $D^0\to \pi^+ \pi^-$.  
We have studied the hypothesis that the measured asymmetries are instead dominated by new physics 
(NP) assuming that $\rm{SU}(3)_{\rm{F}}$ works. To test this hypothesis we invoked two scenarios 
characterized by the U-spin quantum number of the NP interaction. 
We have derived U-spin sum rules between different CP asymmetries which can discriminate between our $\Delta U=0$ and $|\Delta U|=1$ scenarios, for both $D\to PP$ and $D\to PV$ decays. The second scenario is qualitatively different from the SM case; we find six $|\Delta U|=1$ sum rules for   $D\to PP$
and eight ones for $D\to PV$ decays. This large number of experimentally testable relations 
will help to discriminate between NP effects and a SM explanation invoking the breakdown of U-spin symmetry.
One of our CP sum rules holds for both the $\Delta U=0$ and $|\Delta U|=1$ scenarios. This sum rule could be useful to assess the quality of U-spin symmetry irrespective of the presence of NP.
We have also proposed to form differences $\Delta a_{\rm CP}$ between the CP asymmetries in the SCS 
of interest and those in CF and DCS decays to eliminate experimental production and detection asymmetries.
To test our $\rm{SU}(3)_{\rm{F}}$ sum rules more precise data and new measurements are important \cite{LHCb:2018roe,Belle-II:2018jsg}.

%%%%%%%%%%%%%%%%%%%%%%%%%%%%%%%%%%%%
\section*{Acknowledgements}
%%%%%%%%%%%%%%%%%%%%%%%%%%%%%%%%%%%%
The authors thank S.\,Schacht and M.\,Lang for valuable discussion and comment.
The research in this paper was supported by the BMBF
grant 05H21VKKBA, \emph{Theoretische Studien f\"ur Belle II und LHCb} and JSPS KAKENHI Grant Number 24K22879 and JPJSCCA20200002.

%%%%%%%%%%%%%%%%%%%%%%%%%%%%%%%%%%%%
\begin{widetext}
\appendix
%%%%%%%%%%%%%%%%%%%%%%%%%%%%%%%%%%%%
\section{\texorpdfstring{$\mathcal{O}$}{O}(1) Wigner-Eckart invariants of SCS decays}
\label{Sec:WEI}
%%%%%%%%%%%%%%%%%%%%%%%%%%%%%%%%%%%%

The effective Hamiltonian transforms as the product
\begin{equation}
 \mathbf{\bar{3}} \otimes \mathbf{3} \otimes \mathbf{\bar{3}} = (\mathbf{1}\oplus\mathbf{8})\otimes\mathbf{\bar{3}} = \mathbf{\bar{3}}_1 \oplus \mathbf{\bar{3}}_2 \oplus \mathbf{6} \oplus \mathbf{\bar{15}},
\end{equation}
which can be reduced to a direct sum of irreducible representations \cite{Grossman:2012ry}.
For instance, the $\Delta U=0$ contribution proportional to $(\bar{s}s-\bar{d}d)(\bar{u}c)$ contains no operators of $\mathbf{\bar{3}}$. 
These thus lead to penguin contributions in $(\bar{s}s+\bar{d}d)(\bar{u}c)$.
Thanks to the Wigner-Eckart theorem, we can systematically express the symmetry properties of the final and initial states as well as the Hamilton operator, and reduce the number of free parameters in the hadronic matrix elements.
For $\rm{SU}(3)$ it has a similar structure as for $\rm{SU}(2)$ and it follows
\begin{equation}
 \langle P_1P_2 | \mathcal{H} | D \rangle = \sum_w C_w(D,P_1,P_2) X_w \, ,
\end{equation}
where the CG coefficients $C_w$ are also called Wigner-Eckart invariants and $X_w$ are the reduced matrix elements.
We label the Wigner-Eckart invariants as $w=\,$\wigeckinv{R}{i}, see Ref.\,\cite{Grossman:2012ry} for details, where $\mathbf{R}$ is the generating operator in $\mathcal{H}$ and $i$ labels the $i$th reduced element.
Indices of meson representation are dropped while they are clear from the corresponding decays.
The $D\to PP$ and $D\to PV$ Wigner-Eckart invariants of the SCS decay are summarized in Tabs.\,\ref{table:WEI_PP} and \ref{table:WEI_P1P8} as well as Tabs.\,\ref{table:WEI_PV}, \ref{table:WEI_PV2} and \ref{table:WEI_P1V8} which are taken from Ref.\,\cite{Grossman:2012ry}.

%%%%%%%%%%%%%%%%%%%%%%%%%%%%%%%%%
\begin{table}[h]
\centering
\newcommand{\bhline}[1]{\noalign{\hrule height #1}}
\renewcommand{\arraystretch}{1.2}
   \scalebox{1.1}{
  \begin{tabular}{c| c| c| c| c} 
  \rowcolor{white}
  $PP$ mode & (\wigeckinv{\overline{3}}{1},\,\wigeckinv{\overline{3}}{2}) & (\wigeckinv{6}{1},\,\wigeckinv{\overline{15}}{1},\,\wigeckinv{\overline{15}}{2}) & $\Delta U=0$ sum rule & $\Delta U=1$ sum rule\\  \hline
 $D^0\to K^-K^+$ & $\frac{\lambda_b}{4}(0,1)$ & $\frac{\lambda}{2}(1,2,1)$ & Eq.\,(\ref{Eq:PPU0_1})& Eq.\,(\ref{Eq:PPU1_1})\\  
 $D^0\to \pi^-\pi^+$ & $\frac{\lambda_b}{4}(0,1)$ & $-\frac{\lambda}{2}(1,2,1)$ &Eq.\,(\ref{Eq:PPU0_1})& Eq.\,(\ref{Eq:PPU1_1})\\  
 $D^0\to \pi^0\pi^0$ & $\frac{\lambda_b}{4}(0,1)$ & $\frac{\lambda}{2}(1,2,-1)$ &Eq.\,(\ref{Eq:PPU0_4})& Eq.\,(\ref{Eq:PPU1_3})\\  
 $D^0\to \eta_8\eta_8$ & $-\frac{\lambda_b}{12}(2,-3)$ & $\frac{\lambda}{2}(1,2,-1)$ &Eq.\,(\ref{Eq:PPU0_4})& Eqs.\,(\ref{Eq:PPU1_3},\,\ref{Eq:PPU1_4})\\ 
 $D^0\to \eta_8\pi^0$ & $\frac{\lambda_b}{4\sqrt{3}}(1,0)$ & $\frac{\lambda}{2\sqrt{3}}(1,2,-1)$  &Eq.\,(\ref{Eq:PPU0_4})& Eq.\,(\ref{Eq:PPU1_4})\\  \hline
 $D^+\to \pi^0\pi^+$ & $(0,0)$ & $\frac{\lambda}{\sqrt{2}}(0,0,1)$  &Eq.\,(\ref{Eq:PPU0_3})& Eqs.\,(\ref{Eq:PPU1_5},\,\ref{Eq:PPU1_6})\\  
 $D^+\to \eta_8\pi^+$ & $\frac{\lambda_b}{2\sqrt{6}}(1,0)$ & $\frac{\lambda}{\sqrt{6}}(1,-2,-2)$  & Eq.\,(\ref{Eq:PPU0_5})& Eq.\,(\ref{Eq:PPU1_5})\\ 
 $D^+\to \ov{K}^0K^+$ & $\frac{\lambda_b}{4}(1,0)$ & $\frac{\lambda}{2}(1,-2,1)$  &Eq.\,(\ref{Eq:PPU0_2})& Eq.\,(\ref{Eq:PPU1_2})\\ 
 $D^+_s\to \pi^0K^+$ & $\frac{\lambda_b}{4\sqrt{2}}(1,0)$ & $-\frac{\lambda}{2\sqrt{2}}(1,-2,-1)$  &Eqs.\,(\ref{Eq:PPU0_5},\,\ref{Eq:PPU0_6})& Eqs.\,(\ref{Eq:PPU1_5},\,\ref{Eq:PPU1_6})\\  
 $D^+_s\to K^0\pi^+$ & $\frac{\lambda_b}{4}(1,0)$ & $-\frac{\lambda}{2}(1,-2,1)$ &Eqs.\,(\ref{Eq:PPU0_2},\,\ref{Eq:PPU0_6})& Eqs.\,(\ref{Eq:PPU1_2},\,\ref{Eq:PPU1_6})\\  
 $D^+_s\to \eta_8K^+$ & $-\frac{\lambda_b}{4\sqrt{6}}(1,0)$ & $\frac{\lambda}{2\sqrt{6}}(1,-2,-5)$  &Eqs.\,(\ref{Eq:PPU0_5},\,\ref{Eq:PPU0_6})& Eqs.\,(\ref{Eq:PPU1_5},\,\ref{Eq:PPU1_6})\\  \hline
   \end{tabular}
 }
    \caption{\label{table:WEI_PP} 
    The SCS decay Wigner-Eckart invariants and occurrence of the sum rules for $D_{(s)}\to PP$.
    }
\end{table}
%%%%%%%%%%%%%%%%%%%%%%%%%%%%%%%%%

% These are the singlet invariants for PP
%%%%%%%%%%%%%%%%%%%%%%%%%%%%%%%%%
\begin{table}[h]
\centering
\newcommand{\bhline}[1]{\noalign{\hrule height #1}}
\renewcommand{\arraystretch}{1.2}
   \scalebox{1.1}{
  \begin{tabular}{c| c| c| c| c} 
  \rowcolor{white}
  $PP$ mode & (\wigeckinv{\overline{3}}{1},\,\wigeckinv{\overline{3}}{2}) & (\wigeckinv{6}{1},\,\wigeckinv{\overline{15}}{1}) & $\Delta U=0$ sum rule & $\Delta U=1$ sum rule\\  \hline
 $D^0\to \eta_1\eta_1$ & $\frac{\lambda_b}{4}(1,0)$ & $(0,0)$ & ----- & ----- \\  
 $D^0\to \eta_1\eta_8$ & $\frac{\lambda_b}{8\sqrt{6}}(0,1)$ & $-\frac{\lambda}{2}\sqrt{\frac{3}{2}}(1,1)$ & Eq.\,(\ref{Eq:C1}) & Eq.\,(\ref{Eq:C7}) \\  
 $D^0\to \eta_1\pi^0$ & $\frac{\lambda_b}{8\sqrt{2}}(0,1)$ & $\frac{\lambda}{2\sqrt{2}}(1,1)$ & Eq.\,(\ref{Eq:C1}) & Eq.\,(\ref{Eq:C7}) \\ \hline
 $D^+\to \eta_1\pi^+$ & $\frac{\lambda_b}{8}(0,1)$ & $\frac{\lambda}{2}(1,-1)$  & Eq.\,(\ref{Eq:C2}) & Eq.\,(\ref{Eq:C8}) \\  
 $D^+_s\to \eta_1 K^+$ & $\frac{\lambda_b}{8}(0,1)$ & $\frac{\lambda}{2}(-1,1)$  & Eq.\,(\ref{Eq:C2}) & Eq.\,(\ref{Eq:C8}) \\   \hline
   \end{tabular}
 }
    \caption{\label{table:WEI_P1P8} 
    The SCS decay Wigner-Eckart invariants and occurrence of the sum rules for $D_{(s)}\to PP$ including the pseudoscalar SU(3)$_{\rm{F}}$ singlet $\eta_1$. 
    }
\end{table}
%%%%%%%%%%%%%%%%%%%%%%%%%%%%%%%%%

%%%%%%%%%%%%%%%%%%%%%%%%%%%%%%%%%
\begin{table}[h]
\centering
\newcommand{\bhline}[1]{\noalign{\hrule height #1}}
\renewcommand{\arraystretch}{1.2}
   \scalebox{1.13}{
  \begin{tabular}{c| c| c| c| c} 
  \rowcolor{white}
  $PV$ mode & (\wigeckinv{\overline{3}}{1},\,\wigeckinv{\overline{3}}{2},\,\wigeckinv{\overline{3}}{3}) &  (\wigeckinv{6}{1},\,\wigeckinv{6}{2},\,\wigeckinv{6}{3},\,\wigeckinv{\overline{15}}{1},\,\wigeckinv{\overline{15}}{2},\,\wigeckinv{\overline{15}}{3},\,\wigeckinv{\overline{15}}{4}) & $\Delta U=0$ sum rule &$\Delta U=1$ sum rule\\  \hline
 $D^0\to \eta_8 \omega_8$ & $-\frac{\lambda_b}{48}(4,-6,-1)$ & $\frac{\lambda}{4}(0,1,-1,2,1,1,-1)$&Eq.\,(\ref{Eq:PVU0_6})& Eq.\,(\ref{Eq:PVU1_1})\\ 
 $D^0\to \eta_8 \rho^0$ & $\frac{\lambda_b}{16\sqrt{3}}(2,0,1)$ & $-\frac{\lambda}{4\sqrt{3}}(2,1,3,-2,-5,-1,-1)$&Eqs.\,(\ref{Eq:PVU0_6},\,\ref{Eq:PVU0_9})& -----\\  
 $D^0\to K^0 \overline{K}^{*0}$ & $-\frac{\lambda_b}{8}(1,-1,0)$ & $\frac{\lambda}{2}(1,0,0,0,1,-1,0)$&Eq.\,(\ref{Eq:PVU0_1})& Eq.\,(\ref{Eq:PVU1_2})\\  
 $D^0\to \pi^0 \omega_8$ & $\frac{\lambda_b}{16\sqrt{3}}(2,0,1)$ & $\frac{\lambda}{4\sqrt{3}}(2,3,1,2,-3,1,-3)$ &Eqs.\,(\ref{Eq:PVU0_6},\,\ref{Eq:PVU0_9})& -----\\ 
 $D^0\to \pi^0 \rho^0$ & $\frac{\lambda_b}{16}(0,2,1)$ & $-\frac{\lambda}{4}(0,1,-1,2,1,1,-1)$ &Eq.\,(\ref{Eq:PVU0_6})& Eq.\,(\ref{Eq:PVU1_1})\\  
 $D^0\to \ov{K}^0 K^{*0}$ & $-\frac{\lambda_b}{8}(1,-1,0)$ & $-\frac{\lambda}{2}(1,0,0,0,1,-1,0)$ &Eq.\,(\ref{Eq:PVU0_1})& Eq.\,(\ref{Eq:PVU1_2})\\ 
 $D^0\to K^-K^{*+}$ & $\frac{\lambda_b}{8}(0,1,0)$ & $\frac{\lambda}{2}(0,1,0,1,1,1,1)$ &Eq.\,(\ref{Eq:PVU0_2})& Eq.\,(\ref{Eq:PVU1_3})\\  
 $D^0\to \pi^-\rho^+$ & $\frac{\lambda_b}{8}(0,1,0)$ & $-\frac{\lambda}{2}(0,1,0,1,1,1,1)$ &Eq.\,(\ref{Eq:PVU0_2})& Eq.\,(\ref{Eq:PVU1_3})\\ 
 $D^0\to K^+K^{*-}$ & $\frac{\lambda_b}{8}(0,1,1)$ & $-\frac{\lambda}{2}(0,0,1,-1,0,0,0)$ &Eq.\,(\ref{Eq:PVU0_3})& Eq.\,(\ref{Eq:PVU1_4})\\  
 $D^0\to \pi^+\rho^-$ & $\frac{\lambda_b}{8}(0,1,1)$ & $\frac{\lambda}{2}(0,0,1,-1,0,0,0)$ &Eq.\,(\ref{Eq:PVU0_3})& Eq.\,(\ref{Eq:PVU1_4})\\  \hline
   \end{tabular}
   }
    \caption{\label{table:WEI_PV}
   The SCS decay Wigner-Eckart invariants and occurrence of the sum rules for $D^0\to PV$.
    }
\end{table}
%%%%%%%%%%%%%%%%%%%%%%%%%%%%%%%%%

%%%%%%%%%%%%%%%%%%%%%%%%%%%%%%%%%
\begin{table}[h]
\centering
\newcommand{\bhline}[1]{\noalign{\hrule height #1}}
\renewcommand{\arraystretch}{1.2}
   \scalebox{1.13}{
  \begin{tabular}{c |c |c |c |c} 
  \rowcolor{white}
  $PV$ mode & (\wigeckinv{\overline{3}}{1},\,\wigeckinv{\overline{3}}{2},\,\wigeckinv{\overline{3}}{3}) &  (\wigeckinv{6}{1},\,\wigeckinv{6}{2},\,\wigeckinv{6}{3},\,\wigeckinv{\overline{15}}{1},\,\wigeckinv{\overline{15}}{2},\,\wigeckinv{\overline{15}}{3},\,\wigeckinv{\overline{15}}{4}) & $\Delta U=0$ sum rule &$\Delta U=1$ sum rule\\  \hline
 $D^+\to \eta_8\rho^+$ & $\frac{\lambda_b}{8\sqrt{6}}(2,0,1)$ & $-\frac{\lambda}{2\sqrt{6}}(2,1,3,2,-1,1,1)$ &Eqs.\,(\ref{Eq:PVU0_7},\,\ref{Eq:PVU0_9})& Eq.\,(\ref{Eq:PVU1_7})\\  
 $D^+\to \pi^0\rho^+$ & $\frac{\lambda_b}{8\sqrt{2}}(0,0,1)$ & $\frac{\lambda}{2\sqrt{2}}(0,1,1,0,1,-1,1)$ &Eqs.\,(\ref{Eq:PVU0_7},\,\ref{Eq:PVU0_9})& Eq.\,(\ref{Eq:PVU1_7})\\  
 $D^+\to \ov{K}^0 K^{*+}$ & $\frac{\lambda_b}{8}(1,0,0)$ & $\frac{\lambda}{2}(1,1,0,-1,0,0,1)$ &Eqs.\,(\ref{Eq:PVU0_4},\,\ref{Eq:PVU0_9})& Eq.\,(\ref{Eq:PVU1_5})\\  
 $D^+\to K^+\ov{K}^{*0}$ & $\frac{\lambda_b}{8}(1,0,1)$ & $-\frac{\lambda}{2}(1,0,1,1,1,1,0)$ &Eqs.\,(\ref{Eq:PVU0_5},\,\ref{Eq:PVU0_9})& Eq.\,(\ref{Eq:PVU1_6})\\  
 $D^+\to \pi^+\omega_8$ & $\frac{\lambda_b}{8\sqrt{6}}(2,0,1)$ & $\frac{\lambda}{2\sqrt{6}}(2,3,1,-2,-3,-1,-3)$&Eqs.\,(\ref{Eq:PVU0_8},\,\ref{Eq:PVU0_9})& Eq.\,(\ref{Eq:PVU1_8})\\  
 $D^+\to \pi^+\rho^0$ & $-\frac{\lambda_b}{8\sqrt{2}}(0,0,1)$ & $-\frac{\lambda}{2\sqrt{2}}(0,1,1,0,1,-1,-1)$ &Eqs.\,(\ref{Eq:PVU0_8},\,\ref{Eq:PVU0_9})& Eq.\,(\ref{Eq:PVU1_8})\\  \hline
 $D^+_s\to \eta_8 K^{*+}$ & $-\frac{\lambda_b}{8\sqrt{6}}(1,0,-1)$ & $-\frac{\lambda}{2\sqrt{6}}(1,2,3,1,1,-1,2)$ &Eqs.\,(\ref{Eq:PVU0_7},\,\ref{Eq:PVU0_9})& Eq.\,(\ref{Eq:PVU1_7})\\ 
 $D^+_s\to K^0\rho^+$ & $\frac{\lambda_b}{8}(1,0,0)$ & $-\frac{\lambda}{2}(1,1,0,-1,0,0,1)$ &Eq.\,(\ref{Eq:PVU0_4})& Eq.\,(\ref{Eq:PVU1_5})\\  
 $D^+_s\to \pi^0K^{*+}$ & $\frac{\lambda_b}{8\sqrt{2}}(1,0,1)$ & $\frac{\lambda}{2\sqrt{2}}(1,0,1,1,-1,1,0)$ &Eq.\,(\ref{Eq:PVU0_7})& Eq.\,(\ref{Eq:PVU1_7})\\  
 $D^+_s\to K^+\omega_8$ & $-\frac{\lambda_b}{8\sqrt{6}}(1,0,2)$ & $\frac{\lambda}{2\sqrt{6}}(1,3,2,-1,0,-2,-3)$ &Eq.\,(\ref{Eq:PVU0_8})& Eq.\,(\ref{Eq:PVU1_8})\\  
 $D^+_s\to K^+ \rho^0$ & $\frac{\lambda_b}{8\sqrt{2}}(1,0,0)$ & $-\frac{\lambda}{2\sqrt{2}}(1,1,0,-1,-2,0,-1)$ &Eqs.\,(\ref{Eq:PVU0_8},\,\ref{Eq:PVU0_9})& Eq.\,(\ref{Eq:PVU1_8})\\  
 $D^+_s\to \pi^+ K^{*0}$ & $\frac{\lambda_b}{8}(1,0,1)$ & $\frac{\lambda}{2}(1,0,1,1,1,1,0)$&Eq.\,(\ref{Eq:PVU0_5})& Eq.\,(\ref{Eq:PVU1_6})\\  \hline
   \end{tabular}
   }
    \caption{\label{table:WEI_PV2}
    The SCS decay Wigner-Eckart invariants and occurrence of the sum rules for $D^+_{(s)}\to PV$.
    }
\end{table}
%%%%%%%%%%%%%%%%%%%%%%%%%%%%%%%%%

% These are the singlet invariants fro PV
%%%%%%%%%%%%%%%%%%%%%%%%%%%%%%%%%
\begin{table}[h]
\centering
\newcommand{\bhline}[1]{\noalign{\hrule height #1}}
\renewcommand{\arraystretch}{1.2}
   \scalebox{1.13}{
  \begin{tabular}{c| c| c| c| c} 
  \rowcolor{white}
  $PV$ mode & (\wigeckinv{\overline{3}}{1},\,\wigeckinv{\overline{3}}{2},\,\wigeckinv{\overline{3}}{3}) &  (\wigeckinv{6}{1},\,\wigeckinv{6}{2},\,\wigeckinv{\overline{15}}{1},\,\wigeckinv{\overline{15}}{2}) & $\Delta U=0$ sum rule &$\Delta U=1$ sum rule\\  \hline
 $D^0\to \eta_1 \omega_1$ & $\frac{\lambda_b}{8}(1,0,0)$ & $(0,0,0,0)$& ----- & ----- \\ 
 $D^0\to \eta_1 \omega_8$ & $\frac{\lambda_b}{8\sqrt{6}}(0,1,0)$ & $-\frac{\lambda}{2}\sqrt{\frac{3}{2}}(1,0,1,0)$& Eq.\,(\ref{Eq:C3}) & Eq.\,(\ref{Eq:C9}) \\  
 $D^0\to \eta_1 \rho^0$ & $\frac{\lambda_b}{8\sqrt{2}}(0,1,0)$ & $\frac{\lambda}{2\sqrt{2}}(1,0,1,0)$& Eq.\,(\ref{Eq:C3}) & Eq.\,(\ref{Eq:C9}) \\  
 $D^0\to \eta_8 \omega_1$ & $\frac{\lambda_b}{8\sqrt{6}}(0,0,1)$ & $-\frac{\lambda}{2}\sqrt{\frac{3}{2}}(0,1,0,1)$ & Eq.\,(\ref{Eq:C5}) & Eq.\,(\ref{Eq:C11}) \\ 
 $D^0\to \pi^0 \omega_1$ & $\frac{\lambda_b}{8\sqrt{2}}(0,0,1)$ & $\frac{\lambda}{2\sqrt{2}}(0,1,0,1)$ & Eq.\,(\ref{Eq:C5}) & Eq.\,(\ref{Eq:C11}) \\  \hline
 $D^+\to \eta_1\rho^+$ & $\frac{\lambda_b}{8}(0,1,0)$ & $\frac{\lambda}{2}(1,0,-1,0)$ & Eq.\,(\ref{Eq:C4}) & Eq.\,(\ref{Eq:C10}) \\ 
 $D^+\to \pi^+\omega_1$ & $\frac{\lambda_b}{8}(0,0,1)$ & $\frac{\lambda}{2}(0,1,0,-1)$ & Eq.\,(\ref{Eq:C6}) & Eq.\,(\ref{Eq:C12}) \\  
 $D^+_s\to \eta_1K^{*+}$ & $\frac{\lambda_b}{8}(0,1,0)$ & $\frac{\lambda}{2}(-1,0,1,0)$ & Eq.\,(\ref{Eq:C4}) & Eq.\,(\ref{Eq:C10}) \\ 
 $D^+_s\to K^+\omega_1$ & $\frac{\lambda_b}{8}(0,0,1)$ & $\frac{\lambda}{2}(0,-1,0,1)$ & Eq.\,(\ref{Eq:C6}) & Eq.\,(\ref{Eq:C12}) \\ \hline
   \end{tabular}
   }
    \caption{\label{table:WEI_P1V8}
   The SCS decay Wigner-Eckart invariants and occurrence of the sum rules for $D^0\to PV$ including the pseudoscalar and vector SU(3)$_{\rm{F}}$ singlets $\eta_1$ and $\omega_1$.
    }
\end{table}
%%%%%%%%%%%%%%%%%%%%%%%%%%%%%%%%%

%%%%%%%%%%%%%%%%%%%%%%%%%%%%%%%%%%%%
\section{Two-mode sum rules}
\label{Sec:TMSR}
%%%%%%%%%%%%%%%%%%%%%%%%%%%%%%%%%%%%

Thanks to the Wigner-Eckart theorem we can relate the different decay modes based on the group theoretical decomposition and contraction. 
Here we explain the relation between the amplitude sum rule and $a_\text{CP}$ sum rule in the case where only two decay modes are involved.
For the relations involving three or more modes, there is in general no simple formula.
We start from the relations
\begin{align} \label{eq:smallsum}
	P_0^{(P_1P_2)} = c_P\, P_0^{(Q_1Q_2)},\quad A^{(P_1P_2)} = c_A\,A^{(Q_1Q_2)} \, ,
\end{align}
where $c_A$ and $c_P$ are real coefficients which can be read from the tables in Appendix \,\ref{Sec:WEI}. 
We can simply replace the amplitudes $P_0,A_1$ in Eq.\,(\ref{Eq:aCP_U01}) and find the two-modes sum rule 
\begin{align}
	a^{\Delta U = 0}_\text{CP}(P_1P_2) = \frac{c_P}{c_A} \, a^{\Delta U = 0}_\text{CP}(Q_1Q_2) \, .
\end{align}
This can of course also be done for the $|\Delta U| = 1$ case but it is always necessary that both $P$ and $A$ obey sum rules as in Eq.\,(\ref{eq:smallsum}).
For instance sum rules in Eqs.\,(\ref{Eq:PPU0_1},\,\ref{Eq:PPU0_2}) follow from $c_P=1$ and $c_A=-1$ as seen from table \ref{table:WEI_PP}.

%%%%%%%%%%%%%%%%%%%%%%%%%%%%%%%%%%%%
\section{Sum rules for decays into final states with singlet mesons}
\label{Sec:SiSuR}
%%%%%%%%%%%%%%%%%%%%%%%%%%%%%%%%%%%%
So far we discussed the CP asymmetry sum rules with final states of two color octet mesons.
Here we briefly show the SCS sum rules involving color singlet scalar or vector mesons, namely $\eta_1$ or $\omega_1$. These comprise only two-mode sum rules.

For $\Delta U=0$ we have
\begin{align}
    3\,a_\text{CP}^{\Delta U =0}(\eta_1\eta_8) +  a_\text{CP}^{\Delta U =0}(\eta_1\pi^0) = 0,\label{Eq:C1}\\
    a_\text{CP}^{\Delta U =0}(\eta_1K^+) + a_\text{CP}^{\Delta U =0}(\eta_1\pi^+) = 0,\label{Eq:C2}
\end{align}
for $D \rightarrow PP$ and
\begin{align}
    3\,a_\text{CP}^{\Delta U =0}(\eta_1\omega_8) + a_\text{CP}^{\Delta U =0}(\eta_1\rho^0) = 0,\label{Eq:C3}\\
    a_\text{CP}^{\Delta U =0}(\eta_1K^{*+}) + a_\text{CP}^{\Delta U =0}(\eta_1\rho^+) = 0,\label{Eq:C4}\\
    3\,a_\text{CP}^{\Delta U =0}(\eta_8\omega_1) + a_\text{CP}^{\Delta U =0}(\pi^0\omega_1) = 0,\label{Eq:C5}\\
    a_\text{CP}^{\Delta U =0}( K^+\omega_1) + a_\text{CP}^{\Delta U =0}(\pi^+\omega_1) = 0,\label{Eq:C6}
\end{align}
for $D \rightarrow PV$. 

For $\Delta U=1$ we have
\begin{align}
    a_\text{CP}^{\Delta U =1}(\eta_1\eta_8) - a_\text{CP}^{\Delta U =1}(\eta_1\pi^0) = 0 ,\label{Eq:C7}\\
    a_\text{CP}^{\Delta U =1}(\eta_1K^+) -a_\text{CP}^{\Delta U =1}(\eta_1\pi^+) = 0,\label{Eq:C8}
\end{align}
for $D \rightarrow PP$ and
\begin{align}
    a_\text{CP}^{\Delta U =1}(\eta_1\omega_8) - a_\text{CP}^{\Delta U =1}(\eta_1\rho^0) = 0 ,\label{Eq:C9} \\
    a_\text{CP}^{\Delta U =1}(\eta_1K^{*+}) -a_\text{CP}^{\Delta U =1}(\eta_1\rho^+) = 0 ,\label{Eq:C10} \\
    a_\text{CP}^{\Delta U =1}(\eta_8\omega_1) - a_\text{CP}^{\Delta U =1}(\pi^0\omega_1) = 0 ,\label{Eq:C11}\\
    a_\text{CP}^{\Delta U =1}( K^+\omega_1) - a_\text{CP}^{\Delta U =1}(\pi^+\omega_1) = 0,\label{Eq:C12}
\end{align}
for $D \rightarrow PV$.

%%%%%%%%%%%%%%%%%%%%%%%%%%%%%%%%%
\section{Current experimental status and future sensitivity}%
\label{Sec:Input}%
%%%%%%%%%%%%%%%%%%%%%%%%%%%%%%%%%
Tab.\,\ref{tab:CPasym_future} shows the current status and future sensitivity of the CP asymmetries measurements \cite{PDG2022,Belle-II:2018jsg,LHCb:2018roe}.
For the future sensitivity an integrated luminosity of 50\,ab$^{-1}$ and 300\,fb$^{-1}$ is assumed for Belle II and LHCb, respectively.
%%%%%%%%%%%%%%%%%%%%%%%%%%%%%%%%%
\begin{table}[h]
	\centering
	\vspace{0.23cm}
 \renewcommand{\arraystretch}{1.1}
   \scalebox{1.0}{
	\begin{tabular}{c|c|cc |cc|c|c}
		Decay Mode &PDG $a_\text{CP}$ [$\%$]& Belle   & Belle II (50\,ab$^{-1}$) & LHCb & LHCb (300\,fb$^{-1}$)&$\Delta U=0$&$\Delta U=1$\\ \hline
		$D^0\to K^+K^-$ &$-0.07\pm0.11$& $-0.32\pm0.23$ & $\pm0.03$ & $0.077\pm0.057$& $\pm0.007$ &Eq.\,(\ref{Eq:PPU0_1})& Eq.\,(\ref{Eq:PPU1_1})\\ 
		$D^0\to \pi^+\pi^-$ &$0.13\pm0.14$& $0.55\pm0.37$ & $\pm0.05$ & $0.232\pm0.061$ & $\pm0.007$ &Eq.\,(\ref{Eq:PPU0_1})& Eq.\,(\ref{Eq:PPU1_1})\\ 
		$D^0\to \pi^0\pi^0$ &$0.0\pm0.6$& $-0.03\pm0.65$ & $\pm0.09$ &---&---&Eq.\,(\ref{Eq:PPU0_4})& Eq.\,(\ref{Eq:PPU1_3})\\ 
  \hline
		$D^+\to \pi^0 \pi^+$& $0.4\pm1.3$ & $2.31\pm1.26$ & $\pm0.17$ &$-1.3\pm1.1$&---&Eq.\,(\ref{Eq:PPU0_3})& Eqs.\,(\ref{Eq:PPU1_5},\,\ref{Eq:PPU1_6})\\
		$D^+\to \eta \pi^+$& $0.3\pm0.8$& $1.74\pm1.15$ & $\pm0.14$ &$-0.2\pm0.9$&---&Eq.\,(\ref{Eq:PPU0_5})& Eq.\,(\ref{Eq:PPU1_5})\\ 
		$D^+\to \eta^\prime \pi^+$& $-0.6\pm0.7$& $-0.12\pm1.13$ & $\pm0.14$ &$-0.61\pm0.9$&---&Eq.\,(\ref{Eq:PPU0_5})& Eq.\,(\ref{Eq:PPU1_5})\\
		$D^+\to K_s^0K^+ $& $-0.01\pm0.07$& $-0.25\pm0.31$ & $\pm0.04$ &$-0.004\pm0.076$&--- &Eq.\,(\ref{Eq:PPU0_2})& Eq.\,(\ref{Eq:PPU1_2})\\  \hline
		$D^+_s\to  K_s^0\pi^+$& $0.20\pm0.18$& $5.45\pm2.52$ & $\pm0.29$ &$0.16\pm0.18$&---&Eqs.\,(\ref{Eq:PPU0_2},\,\ref{Eq:PPU0_5})& Eqs.\,(\ref{Eq:PPU1_2},\,\ref{Eq:PPU1_6})\\ 
		$D^+_s\to  K^+\eta$& $1.8\pm1.9$& $2.1\pm2.1$ & --- &$0.9\pm3.9$&---&Eqs.\,(\ref{Eq:PPU0_5},\,\ref{Eq:PPU0_6})& Eqs.\,(\ref{Eq:PPU1_5},\,\ref{Eq:PPU1_6})\\ 
		$D^+_s\to  K^+\eta\prime$& $6.0\pm18.9$& --- & --- &---&---&Eqs.\,(\ref{Eq:PPU0_5},\,\ref{Eq:PPU0_6})& Eqs.\,(\ref{Eq:PPU1_5},\,\ref{Eq:PPU1_6})\\ \hline
	\end{tabular}
 }
\caption{\label{tab:CPasym_future}
  Current experimental status and future sensitivity taken from Refs.\,\cite{PDG2022,Belle-II:2018jsg,LHCb:2022lry,LHCb:2018roe}.}
\end{table}

\end{widetext}
%%%%%%%%%%%%%%%%%%%%%%%%%%%%%%%%%%%%
%%%%%%%%%%%%%%%%%%%%%%%%%%%%%%%%%%%%

\bibliography{ref}

\providecommand{\href}[2]{#2}\begingroup\raggedright\begin{thebibliography}{10}

\bibitem{LHCb:2019hro}
{\bfseries LHCb} Collaboration, {\em {Observation of CP Violation in Charm
  Decays},} \href{https://dx.doi.org/10.1103/PhysRevLett.122.211803}{Phys.\
  Rev.\  Lett.\  {\bfseries 122} (2019) 211803} {\ttfamily
  [\href{https://arxiv.org/abs/1903.08726}{arXiv:1903.08726}]}.

\bibitem{LHCb:2022lry}
{\bfseries LHCb} Collaboration, {\em {Measurement of the Time-Integrated CP
  Asymmetry in D0\textrightarrow{}K-K+ Decays},}
  \href{https://dx.doi.org/10.1103/PhysRevLett.131.091802}{Phys.\  Rev.\
  Lett.\  {\bfseries 131} (2023) 091802} {\ttfamily
  [\href{https://arxiv.org/abs/2209.03179}{arXiv:2209.03179}]}.

\bibitem{Cabibbo:1963yz}
N.~Cabibbo, {\em {Unitary Symmetry and Leptonic Decays},}
  \href{https://dx.doi.org/10.1103/PhysRevLett.10.531}{Phys.\  Rev.\  Lett.\
  {\bfseries 10} (1963) 531--533}.

\bibitem{Kobayashi:1973fv}
M.~Kobayashi and T.~Maskawa, {\em {CP Violation in the Renormalizable Theory of
  Weak Interaction},} \href{https://dx.doi.org/10.1143/PTP.49.652}{Prog.\
  Theor.\  Phys.\  {\bfseries 49} (1973) 652--657}.

\bibitem{Khodjamirian:2017zdu}
A.~Khodjamirian and A.~A.~Petrov, {\em {Direct CP asymmetry in $D\to
  \pi^-\pi^+$ and $D\to K^-K^+$ in QCD-based approach},}
  \href{https://dx.doi.org/10.1016/j.physletb.2017.09.070}{Phys.\  Lett.\  B
  {\bfseries 774} (2017) 235--242} {\ttfamily
  [\href{https://arxiv.org/abs/1706.07780}{arXiv:1706.07780}]}.

\bibitem{Chala:2019fdb}
M.~Chala, A.~Lenz, A.~V.~Rusov, and J.~Scholtz, {\em {$\Delta A_{CP}$ within
  the Standard Model and beyond},}
  \href{https://dx.doi.org/10.1007/JHEP07(2019)161}{JHEP {\bfseries 07} (2019)
  161} {\ttfamily [\href{https://arxiv.org/abs/1903.10490}{arXiv:1903.10490}]}.

\bibitem{LHCb:2011osy}
{\bfseries LHCb} Collaboration, {\em {Evidence for CP violation in
  time-integrated $D^0 \to h^-h^+$ decay rates},}
  \href{https://dx.doi.org/10.1103/PhysRevLett.108.111602}{Phys.\  Rev.\
  Lett.\  {\bfseries 108} (2012) 111602} {\ttfamily
  [\href{https://arxiv.org/abs/1112.0938}{arXiv:1112.0938}]}.

\bibitem{Grossman:2006jg}
Y.~Grossman, A.~L.~Kagan, and Y.~Nir, {\em {New physics and CP violation in
  singly Cabibbo suppressed D decays},}
  \href{https://dx.doi.org/10.1103/PhysRevD.75.036008}{Phys.\  Rev.\  D
  {\bfseries 75} (2007) 036008} {\ttfamily
  [\href{https://arxiv.org/abs/hep-ph/0609178}{hep-ph/0609178}]}.

\bibitem{Isidori:2011qw}
G.~Isidori, J.~F.~Kamenik, Z.~Ligeti, and G.~Perez, {\em {Implications of the
  LHCb Evidence for Charm CP Violation},}
  \href{https://dx.doi.org/10.1016/j.physletb.2012.03.046}{Phys.\  Lett.\  B
  {\bfseries 711} (2012) 46--51} {\ttfamily
  [\href{https://arxiv.org/abs/1111.4987}{arXiv:1111.4987}]}.

\bibitem{Grossman:2012eb}
Y.~Grossman, A.~L.~Kagan, and J.~Zupan, {\em {Testing for new physics in singly
  Cabibbo suppressed D decays},}
  \href{https://dx.doi.org/10.1103/PhysRevD.85.114036}{Phys.\ Rev.\  {\bfseries
  D85} (2012) 114036}
{\ttfamily [\href{https://arxiv.org/abs/1204.3557}{arXiv:1204.3557}]}.
%%CITATION = ARXIV:1204.3557;%%.

\bibitem{Giudice:2012qq}
G.~F.~Giudice, G.~Isidori, and P.~Paradisi, {\em {Direct CP violation in charm
  and flavor mixing beyond the SM},}
  \href{https://dx.doi.org/10.1007/JHEP04(2012)060}{JHEP {\bfseries 04} (2012)
  060} {\ttfamily [\href{https://arxiv.org/abs/1201.6204}{arXiv:1201.6204}]}.

\bibitem{Hiller:2012wf}
G.~Hiller, Y.~Hochberg, and Y.~Nir, {\em {Supersymmetric $\Delta A_{CP}$},}
  \href{https://dx.doi.org/10.1103/PhysRevD.85.116008}{Phys.\  Rev.\  D
  {\bfseries 85} (2012) 116008} {\ttfamily
  [\href{https://arxiv.org/abs/1204.1046}{arXiv:1204.1046}]}.

\bibitem{Altmannshofer:2012ur}
W.~Altmannshofer, R.~Primulando, C.-T.~Yu, and F.~Yu, {\em {New Physics Models
  of Direct CP Violation in Charm Decays},}
  \href{https://dx.doi.org/10.1007/JHEP04(2012)049}{JHEP {\bfseries 04} (2012)
  049} {\ttfamily [\href{https://arxiv.org/abs/1202.2866}{arXiv:1202.2866}]}.

\bibitem{Chen:2012usa}
C.-H.~Chen, C.-Q.~Geng, and W.~Wang, {\em {Direct CP Violation in Charm Decays
  due to Left-Right Mixing},}
  \href{https://dx.doi.org/10.1016/j.physletb.2012.11.014}{Phys.\  Lett.\  B
  {\bfseries 718} (2013) 946--950} {\ttfamily
  [\href{https://arxiv.org/abs/1206.5158}{arXiv:1206.5158}]}.

\bibitem{Keren-Zur:2012buf}
B.~Keren-Zur, {\em et al.}, {\em {On Partial Compositeness and the CP asymmetry
  in charm decays},}
  \href{https://dx.doi.org/10.1016/j.nuclphysb.2012.10.012}{Nucl.\  Phys.\  B
  {\bfseries 867} (2013) 394--428} {\ttfamily
  [\href{https://arxiv.org/abs/1205.5803}{arXiv:1205.5803}]}.

\bibitem{Delaunay:2012cz}
C.~Delaunay, J.~F.~Kamenik, G.~Perez, and L.~Randall, {\em {Charming CP
  Violation and Dipole Operators from RS Flavor Anarchy},}
  \href{https://dx.doi.org/10.1007/JHEP01(2013)027}{JHEP {\bfseries 01} (2013)
  027} {\ttfamily [\href{https://arxiv.org/abs/1207.0474}{arXiv:1207.0474}]}.

\bibitem{Dery:2019ysp}
A.~Dery and Y.~Nir, {\em {Implications of the LHCb discovery of CP violation in
  charm decays},} \href{https://dx.doi.org/10.1007/JHEP12(2019)104}{JHEP
  {\bfseries 12} (2019) 104} {\ttfamily
  [\href{https://arxiv.org/abs/1909.11242}{arXiv:1909.11242}]}.

\bibitem{Calibbi:2019bay}
L.~Calibbi, T.~Li, Y.~Li, and B.~Zhu, {\em {Simple model for large CP violation
  in charm decays, $B$-physics anomalies, muon $g-2$ and dark matter},}
  \href{https://dx.doi.org/10.1007/JHEP10(2020)070}{JHEP {\bfseries 10} (2020)
  070} {\ttfamily [\href{https://arxiv.org/abs/1912.02676}{arXiv:1912.02676}]}.

\bibitem{Bause:2020obd}
R.~Bause, H.~Gisbert, M.~Golz, and G.~Hiller, {\em {Exploiting $CP$-asymmetries
  in rare charm decays},}
  \href{https://dx.doi.org/10.1103/PhysRevD.101.115006}{Phys.\  Rev.\  D
  {\bfseries 101} (2020) 115006} {\ttfamily
  [\href{https://arxiv.org/abs/2004.01206}{arXiv:2004.01206}]}.

\bibitem{Buras:2021rdg}
A.~J.~Buras, P.~Colangelo, F.~De~Fazio, and F.~Loparco, {\em {The charm of
  331},} \href{https://dx.doi.org/10.1007/JHEP10(2021)021}{JHEP {\bfseries 10}
  (2021) 021} {\ttfamily
  [\href{https://arxiv.org/abs/2107.10866}{arXiv:2107.10866}]}.

\bibitem{Bause:2022jes}
R.~Bause, {\em et al.}, {\em {U-spin-CP anomaly in charm},}
  \href{https://dx.doi.org/10.1103/PhysRevD.108.035005}{Phys.\  Rev.\  D
  {\bfseries 108} (2023) 035005} {\ttfamily
  [\href{https://arxiv.org/abs/2210.16330}{arXiv:2210.16330}]}.

\bibitem{Brod:2012ud}
J.~Brod, Y.~Grossman, A.~L.~Kagan, and J.~Zupan, {\em {A Consistent Picture for
  Large Penguins in $D \to\pi^+ \pi^-, K^+ K^-$},}
  \href{https://dx.doi.org/10.1007/JHEP10(2012)161}{JHEP {\bfseries 10} (2012)
  161} {\ttfamily [\href{https://arxiv.org/abs/1203.6659}{arXiv:1203.6659}]}.

\bibitem{Grossman:2019xcj}
Y.~Grossman and S.~Schacht, {\em {The emergence of the $\Delta U=0$ rule in
  charm physics},} \href{https://dx.doi.org/10.1007/JHEP07(2019)020}{JHEP
  {\bfseries 07} (2019) 020} {\ttfamily
  [\href{https://arxiv.org/abs/1903.10952}{arXiv:1903.10952}]}.

\bibitem{Soni:2019xko}
A.~Soni, {\em {Resonance enhancement of Charm CP}.} {\ttfamily
  \href{https://arxiv.org/abs/1905.00907}{arXiv:1905.00907}}.

\bibitem{Schacht:2021jaz}
S.~Schacht and A.~Soni, {\em {Enhancement of charm CP violation due to nearby
  resonances},} \href{https://dx.doi.org/10.1016/j.physletb.2021.136855}{Phys.\
   Lett.\  B {\bfseries 825} (2022) 136855} {\ttfamily
  [\href{https://arxiv.org/abs/2110.07619}{arXiv:2110.07619}]}.

\bibitem{Pirtskhalava:2011va}
D.~Pirtskhalava and P.~Uttayarat, {\em {CP Violation and Flavor SU(3) Breaking
  in D-meson Decays},}
  \href{https://dx.doi.org/10.1016/j.physletb.2012.04.039}{Phys.\ Lett.\
  {\bfseries B712} (2012) 81--86}
{\ttfamily [\href{https://arxiv.org/abs/1112.5451}{arXiv:1112.5451}]}.
%%CITATION = ARXIV:1112.5451;%%.

\bibitem{Bhattacharya:2012ah}
B.~Bhattacharya, M.~Gronau, and J.~L.~Rosner, {\em {CP asymmetries in
  singly-Cabibbo-suppressed $D$ decays to two pseudoscalar mesons},}
  \href{https://dx.doi.org/10.1103/PhysRevD.85.054014}{Phys.\  Rev.\  D
  {\bfseries 85} (2012) 054014} {\ttfamily
  [\href{https://arxiv.org/abs/1201.2351}{arXiv:1201.2351}]}.

\bibitem{Feldmann:2012js}
T.~Feldmann, S.~Nandi, and A.~Soni, {\em {Repercussions of Flavour Symmetry
  Breaking on CP Violation in D-Meson Decays},}
  \href{https://dx.doi.org/10.1007/JHEP06(2012)007}{JHEP {\bfseries 06} (2012)
  007} {\ttfamily [\href{https://arxiv.org/abs/1202.3795}{arXiv:1202.3795}]}.

\bibitem{Bhattacharya:2012kq}
B.~Bhattacharya, M.~Gronau, and J.~L.~Rosner, {\em {Direct CP Violation in D
  Decays in view of LHCb and CDF Results}.}
{\ttfamily \href{https://arxiv.org/abs/1207.0761}{arXiv:1207.0761}}.
%%CITATION = ARXIV:1207.0761;%%.

\bibitem{Grossman:2012ry}
Y.~Grossman and D.~J.~Robinson, {\em {SU(3) Sum Rules for Charm Decay},}
  \href{https://dx.doi.org/10.1007/JHEP04(2013)067}{JHEP {\bfseries 04} (2013)
  067} {\ttfamily [\href{https://arxiv.org/abs/1211.3361}{arXiv:1211.3361}]}.

\bibitem{Hiller:2012xm}
G.~Hiller, M.~Jung, and S.~Schacht, {\em {SU(3)-flavor anatomy of nonleptonic
  charm decays},} \href{https://dx.doi.org/10.1103/PhysRevD.87.014024}{Phys.\
  Rev.\  D {\bfseries 87} (2013) 014024} {\ttfamily
  [\href{https://arxiv.org/abs/1211.3734}{arXiv:1211.3734}]}.

\bibitem{Grossman:2013lya}
Y.~Grossman, Z.~Ligeti, and D.~J.~Robinson, {\em {More Flavor SU(3) Tests for
  New Physics in CP Violating B Decays},}
  \href{https://dx.doi.org/10.1007/JHEP01(2014)066}{JHEP {\bfseries 01} (2014)
  066} {\ttfamily [\href{https://arxiv.org/abs/1308.4143}{arXiv:1308.4143}]}.

\bibitem{Hiller:2013awa}
G.~Hiller, M.~Jung, and S.~Schacht, {\em {SU(3)$_{F}$ in nonleptonic charm
  decays},} \href{https://dx.doi.org/10.22323/1.180.0371}{PoS {\bfseries
  EPS-HEP2013} (2013) 371} {\ttfamily
  [\href{https://arxiv.org/abs/1311.3883}{arXiv:1311.3883}]}.

\bibitem{Muller:2015lua}
S.~M\"uller, U.~Nierste, and S.~Schacht, {\em {Topological amplitudes in $D$
  decays to two pseudoscalars: A global analysis with linear $SU(3)_F$
  breaking},} \href{https://dx.doi.org/10.1103/PhysRevD.92.014004}{Phys.\
  Rev.\  D {\bfseries 92} (2015) 014004} {\ttfamily
  [\href{https://arxiv.org/abs/1503.06759}{arXiv:1503.06759}]}.

\bibitem{Muller:2015rna}
S.~M\"uller, U.~Nierste, and S.~Schacht, {\em {Sum Rules of Charm CP
  Asymmetries beyond the SU(3)$_F$ Limit},}
  \href{https://dx.doi.org/10.1103/PhysRevLett.115.251802}{Phys.\  Rev.\
  Lett.\  {\bfseries 115} (2015) 251802} {\ttfamily
  [\href{https://arxiv.org/abs/1506.04121}{arXiv:1506.04121}]}.

\bibitem{Gavrilova:2022hbx}
M.~Gavrilova, Y.~Grossman, and S.~Schacht, {\em {The mathematical structure of
  U-spin amplitude sum rules},}
  \href{https://dx.doi.org/10.1007/JHEP08(2022)278}{JHEP {\bfseries 08} (2022)
  278} {\ttfamily [\href{https://arxiv.org/abs/2205.12975}{arXiv:2205.12975}]}.

\bibitem{Schacht:2022kuj}
S.~Schacht, {\em {A U-spin anomaly in charm CP violation},}
  \href{https://dx.doi.org/10.1007/JHEP03(2023)205}{JHEP {\bfseries 03} (2023)
  205} {\ttfamily [\href{https://arxiv.org/abs/2207.08539}{arXiv:2207.08539}]}.

\bibitem{Nierste:2017cua}
U.~Nierste and S.~Schacht, {\em {Neutral $D\rightarrow K K^*$ decays as
  discovery channels for charm CP violation},}
  \href{https://dx.doi.org/10.1103/PhysRevLett.119.251801}{Phys.\  Rev.\
  Lett.\  {\bfseries 119} (2017) 251801} {\ttfamily
  [\href{https://arxiv.org/abs/1708.03572}{arXiv:1708.03572}]}.

\bibitem{deSwart:1963pdg}
J.~J.~de~Swart, {\em {The Octet model and its Clebsch-Gordan coefficients},}
  \href{https://dx.doi.org/10.1103/RevModPhys.35.916}{Rev.\  Mod.\  Phys.\
  {\bfseries 35} (1963) 916--939}. [Erratum: Rev.Mod.Phys. 37, 326--326
  (1965)].

\bibitem{Kaeding:1995vq}
T.~A.~Kaeding, {\em {Tables of SU(3) isoscalar factors},}
  \href{https://dx.doi.org/10.1006/adnd.1995.1011}{Atom.\  Data Nucl.\  Data
  Tabl.\  {\bfseries 61} (1995) 233--288} {\ttfamily
  [\href{https://arxiv.org/abs/nucl-th/9502037}{nucl-th/9502037}]}.

\bibitem{ATLAS:2017eqx}
{\bfseries ATLAS} Collaboration, {\em {Search for new phenomena in dijet events
  using 37 fb$^{-1}$ of $pp$ collision data collected at $\sqrt{s}=$13 TeV with
  the ATLAS detector},}
  \href{https://dx.doi.org/10.1103/PhysRevD.96.052004}{Phys.\  Rev.\  D
  {\bfseries 96} (2017) 052004} {\ttfamily
  [\href{https://arxiv.org/abs/1703.09127}{arXiv:1703.09127}]}.

\bibitem{ATLAS:2023fod}
{\bfseries ATLAS} Collaboration, {\em {Summary Plots for Heavy Particle
  Searches and Long-lived Particle Searches - March 2023}.}

\bibitem{PDG2022}
{\bfseries Particle Data Group} Collaboration, {\em {Review of Particle
  Physics},} \href{https://dx.doi.org/10.1093/ptep/ptac097}{PTEP {\bfseries
  2022} (2022) 083C01}.

\bibitem{Belle-II:2018jsg}
{\bfseries Belle-II} Collaboration, {\em {The Belle II Physics Book},}
  \href{https://dx.doi.org/10.1093/ptep/ptz106}{PTEP {\bfseries 2019} (2019)
  123C01} {\ttfamily
  [\href{https://arxiv.org/abs/1808.10567}{arXiv:1808.10567}]}. [Erratum: PTEP
  2020, 029201 (2020)].

\bibitem{LHCb:2018roe}
{\bfseries LHCb} Collaboration, {\em {Physics case for an LHCb Upgrade II -
  Opportunities in flavour physics, and beyond, in the HL-LHC era}.} {\ttfamily
  \href{https://arxiv.org/abs/1808.08865}{arXiv:1808.08865}}.

\bibitem{Bergmann:1999pm}
S.~Bergmann and Y.~Nir, {\em {New physics effects in doubly Cabibbo suppressed
  D decays},} \href{https://dx.doi.org/10.1088/1126-6708/1999/09/031}{JHEP
  {\bfseries 09} (1999) 031} {\ttfamily
  [\href{https://arxiv.org/abs/hep-ph/9909391}{hep-ph/9909391}]}.

\bibitem{Bigi:1994aw}
I.~I.~Y.~Bigi and H.~Yamamoto, {\em {Interference between Cabibbo allowed and
  doubly forbidden transitions in D ---\ensuremath{>} K(S), K(L) + pi's
  decays},} \href{https://dx.doi.org/10.1016/0370-2693(95)00285-S}{Phys.\
  Lett.\  B {\bfseries 349} (1995) 363--366} {\ttfamily
  [\href{https://arxiv.org/abs/hep-ph/9502238}{hep-ph/9502238}]}.

\bibitem{Grossman:2011zk}
Y.~Grossman and Y.~Nir, {\em {CP Violation in $\tau^\pm \to \pi^\pm K_S\nu$ and
  $D^\pm \to \pi^\pm K_S$: The Importance of $K_S - K_L$ Interference},}
  \href{https://dx.doi.org/10.1007/JHEP04(2012)002}{JHEP {\bfseries 04} (2012)
  002} {\ttfamily [\href{https://arxiv.org/abs/1110.3790}{arXiv:1110.3790}]}.

\end{thebibliography}\endgroup

\end{document}